%% file: nysa_polana.tex
\documentclass[12pt, preprint]{aastex}

\usepackage{xtab}
\usepackage{amsmath}
\usepackage{graphicx}
\usepackage{subcaption}
\usepackage{booktabs}
\usepackage{enumerate}
\usepackage{natbib}
\usepackage{dcolumn}
\usepackage{float}
\usepackage{siunitx}
\usepackage{multicol}
\usepackage{array}

\include {psfig}

\newcommand{\degree}{\ensuremath{^\circ}}

\newcolumntype{d}[1]{D{.}{.}{#1} }

\journal{Icarus}

\begin{document}

\title{Collisional family structure within the Nysa-Polana complex}

\author{Melissa J$.$ Dykhuis, Richard Greenberg}
\affil{Lunar and Planetary Laboratory, University of Arizona, 1629 E. University Blvd., Tucson, AZ 85719, USA}
\email{dykhuis@lpl.arizona.edu}

\renewcommand{\abstractname}{\newpage ABSTRACT}

\begin {abstract}
The Nysa-Polana complex is a group of low-inclination asteroid families in the inner main belt, bounded in semimajor axis by the Mars-crossing region and the Jupiter 3:1 mean-motion resonance. This group is important as the most likely source region for the target of the OSIRIS-REx mission, (101955) Bennu; however, family membership in the region is complicated by the presence of several dynamically overlapping families with a range of surface reflectance properties. 

The large S-type structure in the region appears to be associated with the parent body (135) Hertha, and displays an ($e_\text{P},a_\text{P}$) correlation consistent with a collision event near true anomaly of $\sim180\degree$ with ejecta velocity $v_\text{ej} \sim 285$~m/s. The ejecta distribution from a collision with these orbital properties is predicted to have a maximum semimajor axis dispersion of $\delta a_{ej} = 0.005 \pm 0.008$~AU, which constitutes only a small fraction (7\%) of the observed semimajor axis dispersion, the rest of which is attributed to the Yarkovsky effect. The age of the family is inferred from the Yarkovsky dispersion to be $300^{+60}_{-50}$~My. 

Objects in a smaller cluster that overlaps the large Hertha family in proper orbital element space have reflectance properties more consistent with the X-type (135) Hertha than the surrounding S-type family. These objects form a distinct Yarkovsky ``V'' signature in ($a_\text{P},H$) space, consistent with a more recent collision, which appears to also be dynamically connected to (135) Hertha. Production of two families with different reflectance properties from a single parent could result from the partial differentiation of the parent, shock darkening effects, or other causes. 

The Nysa-Polana complex also contains a low-albedo family associated with (142) Polana \citep[called ``New Polana'' by][]{walsh2013}, and two other low-albedo families associated with (495) Eulalia. The second Eulalia family may be a high-$a_\text{P}$, low-$e_\text{P}$, low-$i_\text{P}$ component of the first Eulalia family-forming collision, possibly explained by an anisotropic ejection field.
\end {abstract}



\newpage
\section{Introduction}
\label{introduction} 

The low-inclination, inner main belt asteroid population is an important source region for near-Earth objects (NEOs) and terrestrial planet impactors in the past, present and future \citep{scholl1991,morbidelli2003}. This region, which includes the Nysa-Polana complex, poses a challenge for family identification and analysis techniques, because it is dynamically very crowded, with many families overlapping in proper orbital element space. Recent efforts have used newly available color information from the Sloan Digital Sky Survey's Moving Objects Catalog \citep[SDSSMOC\footnote{http://www.astro.washington.edu/users/ivezic/sdssmoc/},][]{ivezic2002} and albedo information from the Wide-field Infrared Survey Explorer \citep[WISE/NEOWISE\footnote{http://irsa.ipac.caltech.edu/Missions/wise.html},][]{wright2010,mainzer2011} to augment the orbital databases in an effort to further distinguish families on the basis of their unique reflectance properties \citep{parker2008,masiero2013,carruba2013,milani2014,walsh2013,dykhuis2014}. 

Analysis of the taxonomically diverse and dynamically overlapping families within the Nysa-Polana complex has suffered from a long and complicated history of identification and re-identification, often with changes in membership, taxonomy, and nomenclature. In order to place the current work in context, we first review the history of these identifications and the rationales that led to each. The various names assigned to each family structure throughout the literature are listed in Table~\ref{table1}. In the interest of connecting our work to future efforts, we also identify the Nysa-Polana region with the Family Identification Number (FIN) of 405, assigned to the region by Nesvorn{\'y} et al. and Masiero et al., both submitted. In this classification scheme, the Hertha1 family would correspond to 405a, the Polana family to 405b, and the Eulalia1 family to 405c. The Hertha2 and Eulalia2 families would await further confirmation from additional data.

The Nysa-Polana complex was first identified as a single family by \citet{brouwer1951}. \citet{zellner1977} subsequently distinguished two family structures in the region, and suggested an origin scenario that attempted to include both of the largest neighboring objects (44) Nysa and (135) Hertha as fragments from the same parent body. \citet{williams1979} confirmed the existence of two families, labeled as ``W-24'' and ``W-160.'' With the addition of reflectance property data from the Eight-Color Asteroid Survey (ECAS), \citet{tedesco1982} discovered several objects in the so-called ``Nysa'' family to be the otherwise rare ``F'' spectral type (using the taxonomy scheme developed by the authors for the ECAS). This spectral dissimilarity led \citet{bell1989} to conclude that the ``Nysa'' family members were unrelated to the E-type (44) Nysa, and should rather be associated with the F-type (142) Polana, labeling (44) Nysa an interloper in that family. \citet{kelley1994} suggested an origin scenario involving partial differentiation in attempt to link the E-type (44) Nysa and M-type (135) Hertha as ``parents'' of the F-type family, based on a shared silicate absorption feature near 0.9~$\mu$m. 

\input{./Tables/summary_table_tolatex.txt}
 {\scriptsize Abbreviations: Ze77 = \citet{zellner1977}; Wi79 = \citet{williams1979}; B89 = \citet{bell1989}; Za95 = \citet{zappala1995}; \\C01 = \citet{cellino2001}; MD05 = \citet{mothe-diniz2005}; D12 = \citet{delbo2012}; Wa13 = \citet{walsh2013}; \\M14 = \citet{milani2014}. }
 \caption{\footnotesize Summary of the various names of Nysa-Polana complex families used in the literature, showing sources that modified the family nomenclature. See the text for a more detailed summary of each literature source. The first column shows the five structures of the Nysa-Polana complex, using the naming conventions of the present work. The spectral types in column 2 are broad categories only, meant to reflect the colors observed by the SDSS; the individual literature sources often use subcategories or parallel categories (such as Xc, Xk, E, F, B, etc.). Ditto marks indicate that the Eulalia1 structure, defined in Section \ref{other_dark}, was typically joined to the Polana structure until its distinction in the work of Wa13. The question mark expresses low confidence reported by Wa13. MD05 lumped the low-$e_\text{P}$ portion of our Hertha1 structure together with our Hertha2 structure, retaining the name Mildred for the higher-$e_\text{P}$ portion of Hertha1. }
 \label{table1}
\end{table}

\citet{zappala1995} used two different clustering techniques to identify family members in the Nysa-Polana region, and found two overlapping families that the clustering methods failed to reliably distinguish. The Polana family was labeled as a possible subsequent collision on a Nysa family member.

\citet{cellino2001} undertook a spectroscopic campaign in an attempt to clarify membership in the region. They found three objects with X-type taxonomy among the F-type Polana objects and S-type Hertha objects. They renamed the Hertha objects as the ``Mildred'' family, based on the dissimilarity between the M-type (135) Hertha and the S-type family previously associated with it. 

Further observations and dynamical analysis by \citet{mothe-diniz2005} and \citet{alvarez-candal2006} identified three more objects as X-types, including (44) Nysa and (135) Hertha. \citet{mothe-diniz2005} separated the Mildred family into two groups based on their eccentricities, and noted that the lower-$e_\text{P}$ group contained a mix of X- and S-type objects, while the higher-$e_\text{P}$ group contained primarily S-type objects only. They combined the lower-$e_\text{P}$ portion of the Mildred family with what we now call the Hertha2 family to produce a ``Hertha'' family, while the higher-$e_\text{P}$ component retained the name ``Mildred.'' In addition, they renamed the Polana family as ``McCuskey.'' More recently, \citet{delbo2012} tested a new classification algorithm on the region, and identified three groups of objects based on reflectance properties, confirming the presence of S-, X- and C-type spectra in the Nysa-Polana region.

\citet{walsh2013} used the recent data available from WISE/NEOWISE to study the dark (B- and C-type) objects in the Nysa-Polana complex, which until then had been considered a single family (usually called ``Polana''). They identified the structures of at least two old families, which they associated with (142) Polana and (495) Eulalia. A third structure at lower eccentricities was identified with (112) Iphigenia, although with low confidence in its significance or its independence from the Eulalia family.

The most recent analyses \citep{masiero2013,carruba2013,milani2014} incorporated major results from large reflectance surveys, primarily WISE and SDSS, and used automated clustering methods applied to this extended parameter space. In general, they confirmed the presence of at least one low-albedo (B- or C-type) collisional family and at least one high-albedo (S-type) collisional family in this region. \citet{milani2014} identified the first cluster in Table~\ref{table1} with a different parent asteroid, (3583) Burdett.

Here, in addition to seeking clusters in the dynamical and reflectance parameter space, we seek signatures of dynamical evolution that distinguish multiple families within a single cluster. Specifically, by considering cluster distributions in $a_\text{P}-H$ space, the distinct ``V-shape'' patterns of Yarkovsky evolution can be recognized. This technique was demonstrated recently for the ``Eulalia'' and ``New Polana'' families by \citet{walsh2013}, and was also successfully applied in our earlier analysis of the large and diffuse Flora family, which dominates the inner main belt \citep{dykhuis2014}. In addition, for the Hertha1 family, we extract collisional parameter information from the slope of the family's $a_\text{P}$-$e_\text{P}$ distribution.

In Section \ref{overview}, we give an overview of the geography of the Nysa-Polana region in proper elements and reflectance properties. Then, we separate the objects on the basis of their $a^*$ color parameter \citep[as defined by][]{ivezic2001}, and study first the objects with $a^* > -0.015$ (Section \ref{high_astar}), which we show to be one of two families likely associated with (135) Hertha (designated the Hertha1 family, see Table~\ref{table1}). Next we study the objects with $a^* < -0.015$ (Section \ref{low_astar}), identifying four structures and their likely parent objects. 

\section{Overview of the region}
\label{overview}

\begin{figure}
\vspace{-0.4in}
\begin{center}
  \begin{subfigure}{\textwidth}
    \begin{centering}
    \includegraphics[width=3.8in]{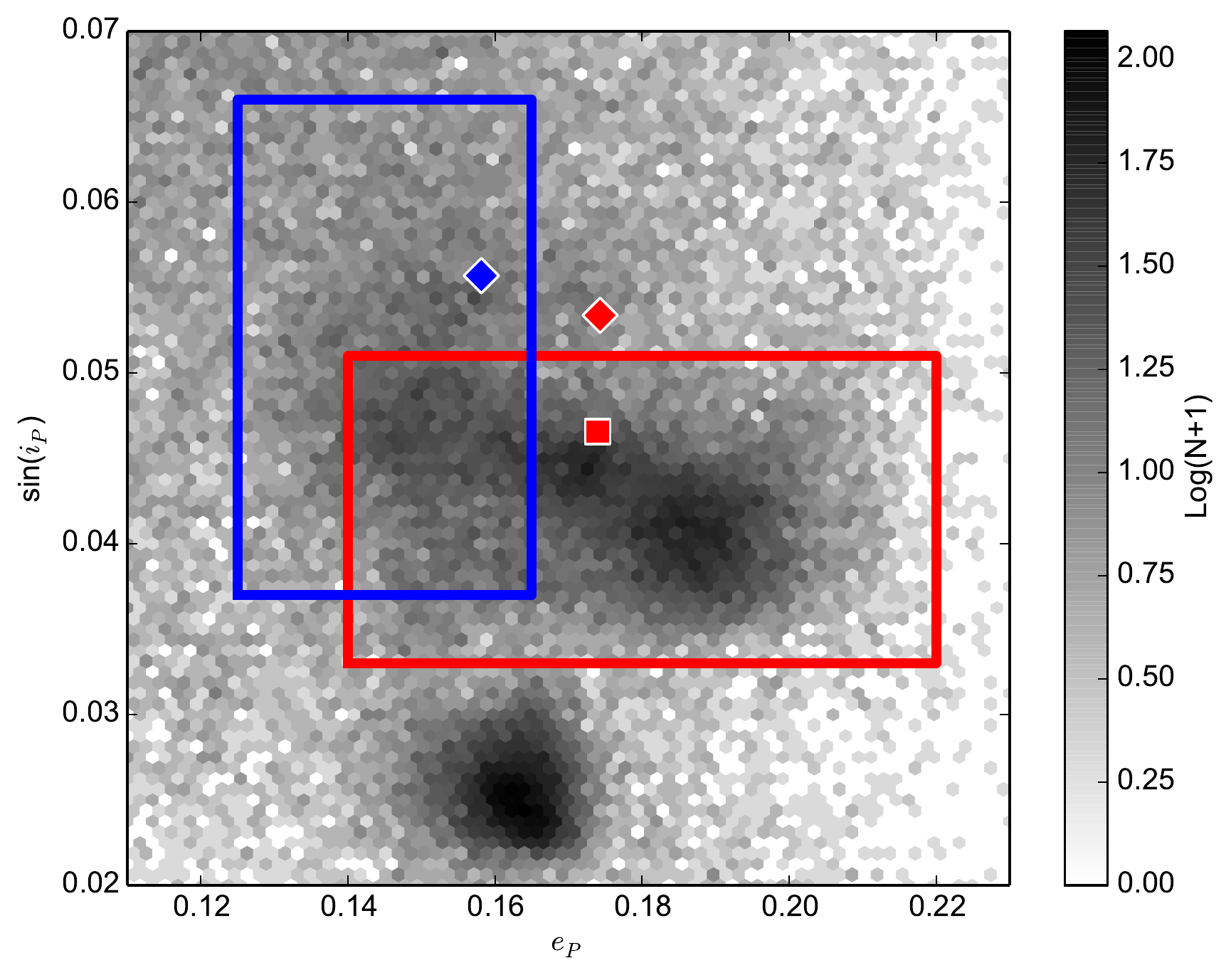}
    \vspace{-0.1in}
    \caption{}
    \label{01_orbital_a}
    \end{centering}
  \end{subfigure}

  \begin{subfigure}{0.49\textwidth}
    \includegraphics[width=3.1in]{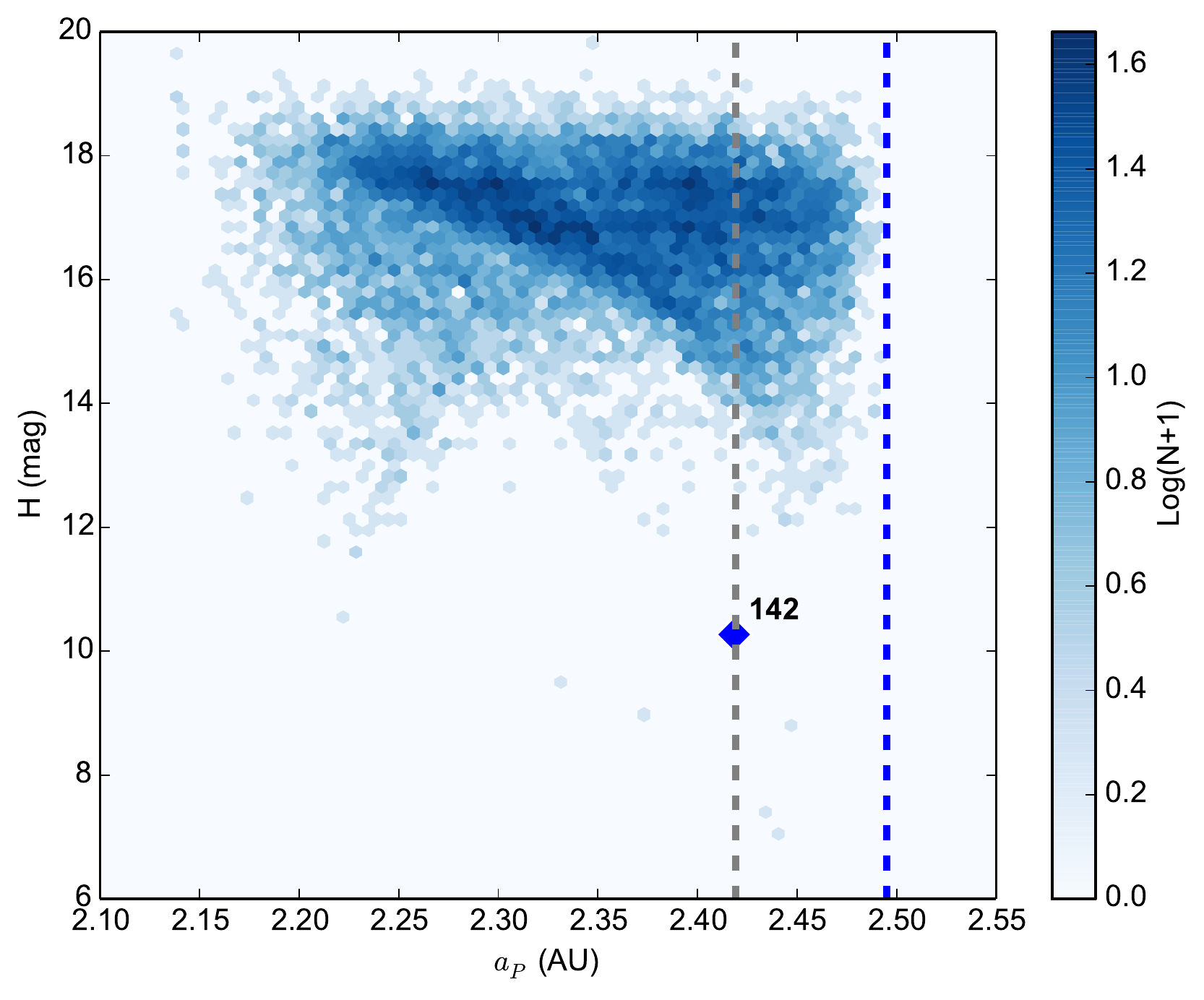}
    \vspace{-0.1in}
    \caption{}
    \label{01_orbital_b}
  \end{subfigure}
  \begin{subfigure}{0.49\textwidth}
    \includegraphics[width=3.1in]{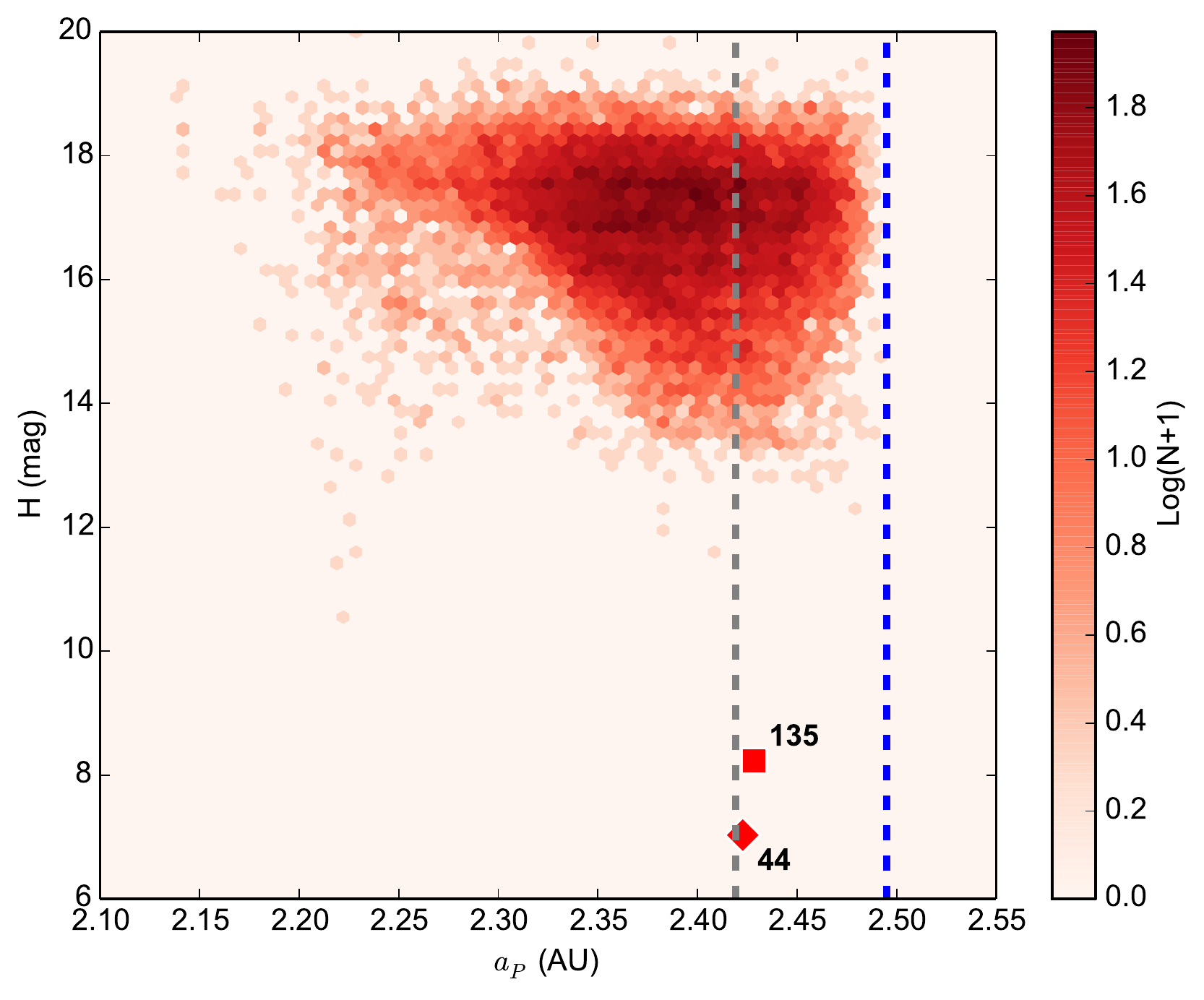}
    \vspace{-0.1in}
    \caption{}
    \label{01_orbital_c}
  \end{subfigure}
  \caption {\footnotesize (a) Density distribution in proper $e_\text{P}$ and $\sin i_\text{P}$ of all objects with well-determined orbits between 2.1 and 2.5~AU. The red and blue boxes indicate the rough boundaries of the regions dominated by S- and C-type objects. The large asteroids (44) Nysa, (135) Hertha and (142) Polana are indicated by the red diamond, red square and blue diamond, respectively. The structure at $e_\text{P} \sim 0.16$ and $\sin i_\text{P} \sim 0.025$ is the Massalia family; the Flora family muddies the upper left half of the plot. Subplots (b) and (c) show the distribution in proper semimajor axis and absolute magnitude of the objects in the blue and red boxes of (a). The gray and blue dotted lines mark the location of the Mars 1:2 and Jupiter 3:1 mean-motion resonances ($a_\text{P}$ = 2.419~AU and 2.495~AU, respectively). Several characteristic Yarkovsky ``V'' patterns appear in plots (b) and (c), suggesting the presence of multiple evolved families. }
\label{01_orbital}
\end{center}
\end{figure}

The orbital distribution of asteroids in the Nysa-Polana region (2.1~AU $< a_\text{P} <$ 2.5~AU, $0.11 < e_\text{P} < 0.22$, and $0.02 < \sin i_\text{P} < 0.07$) with reliable proper orbital element data is shown in Fig.~\ref{01_orbital}. The catalog of synthetic proper elements is the same as that used in \citet{dykhuis2014}, computed using the Orbit9 software available from the Asteroids Dynamic Site (AstDyS)\footnote{http://hamilton.dm.unipi.it/astdys/}. The blue and red boxes in Fig.~\ref{01_orbital_a} roughly represent the regions associated with the low-albedo (C-type) and higher-albedo (S-type) groups, respectively, that have typically been identified in the region. Asteroid (142) Polana lies within the former range, (135) Hertha in the latter, and (44) Nysa in neither.

\begin {center}
EDITOR: PLACE FIGURE \ref{01_orbital} HERE
\end {center}

Figs.~\ref{01_orbital_b} and \ref{01_orbital_c} show the distributions in proper semimajor axis, $a_\text{P}$, vs. absolute magnitude, $H$, of the objects from the sub-regions marked in Fig.~\ref{01_orbital_a}. A ``V'' shape in $a_\text{P}-H$ space is generally attributed to the Yarkovsky effect, which yields a size-dependent spread in semimajor axes for a collisional family. The signatures of several overlapping V's are evident in Figs.~\ref{01_orbital_b} and \ref{01_orbital_c}, suggesting the presence of multiple families. 

The boundary of each V-shaped Yarkovsky envelope can be described by the limiting distance in semimajor axis $\Delta a_\text{P}$ from the family's central semimajor axis location, which is generally assumed to be the proper semimajor axis of the largest remnant of the family. The distance is described by the relation \citep[from][]{vokrouhlicky2006}:

\begin{equation}
 \Delta a_\text{P} = C \cdot 10^{H/5}.
 \label{Cparam}
\end{equation}

\noindent The $C$ parameter (which for convenience we later refer to in units of mAU = $10^{-3}$~AU) is directly related to the family's dynamical evolution due to the Yarkovsky effect. Families with an older age will show more semimajor axis evolution, and reach higher boundary values of $C$ than their younger counterparts; thus the boundary of the $C$ parameter distribution for a given family can be used as a simple means to estimate the age, assuming appropriate calibration of the Yarkovsky semimajor axis drift rates and estimation of the semimajor axis dispersion due to the original collision ejection field.

The largest objects in the Nysa-Polana region --- (44) Nysa, (135) Hertha, and (142) Polana --- all lie quite close in semimajor axis to the Mars 1:2 mean-motion resonance (dotted line in Figs.~\ref{01_orbital_b} and \ref{01_orbital_c}, located at $a_\text{P}$ = 2.419~AU). The effects of this resonance were studied in detail by \citet{gallardo2011}, and must be taken into account in studies of the families in the inner main belt, as discussed in Section \ref{high_astar}.

\begin{figure}
\vspace{-0.3in}
\begin{center}
\includegraphics [width=3.8in]{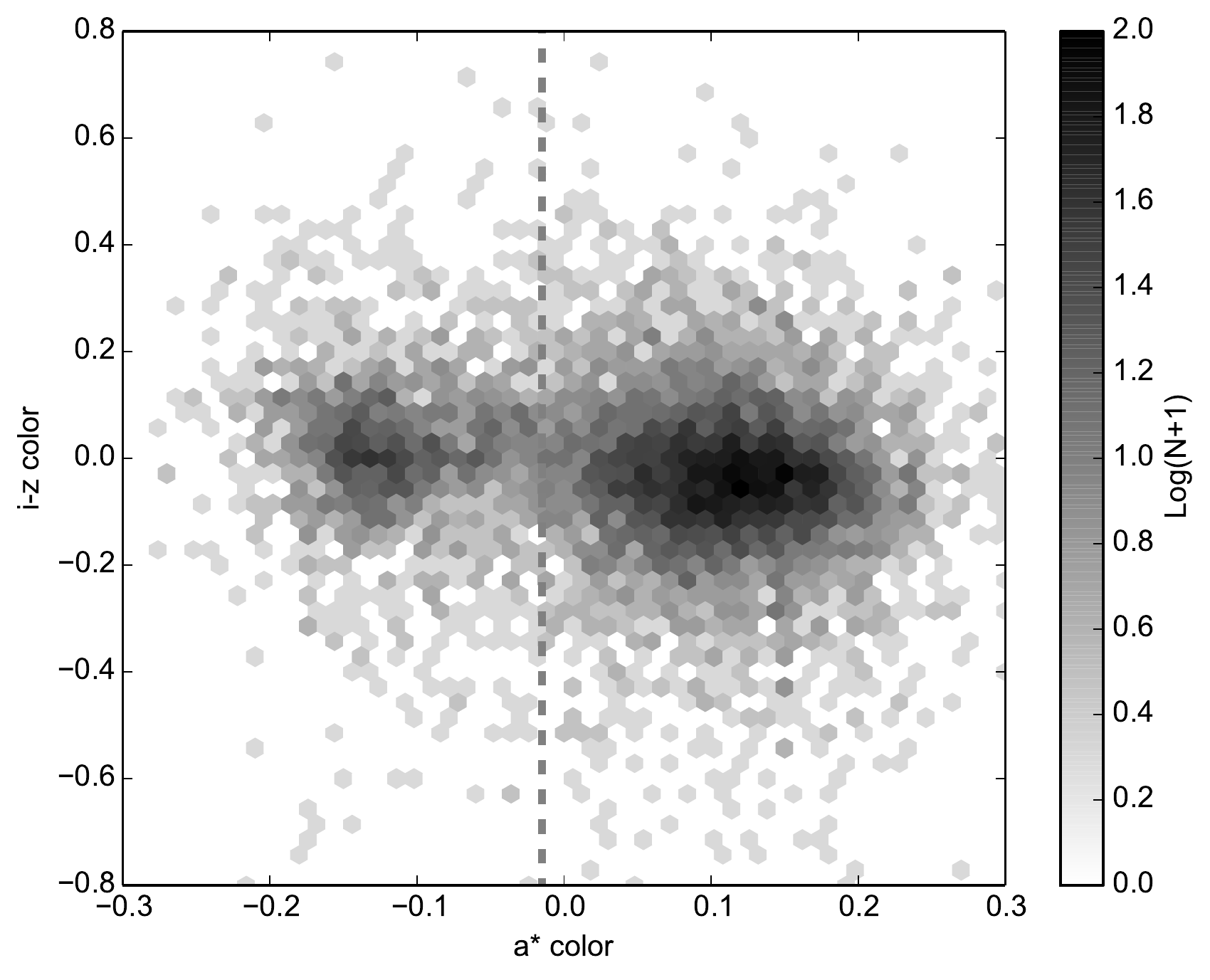}
\caption {\footnotesize Distribution of SDSS colors ($a^*$ and $i-z$) for all bodies included in Fig.~\ref{01_orbital}a for which adequate color data are available. The two major clusters seen here on either side of the boundary at $a^* = -0.015$ (dotted line) are also distinct in the three-dimensional parameter space ($a^*$,$i-z$,$p_\text{V}$). Generally, objects with larger $a^*$ have reflectance properties consistent with S spectral types, while those with smaller $a^*$ are darker spectral types (e.g., C- or X-types).}
\label{02_reflectance}
\end{center}
\end{figure}

Color information for the 8396 Nysa-Polana region objects that have SDSS color data (but not necessarily albedo data) are plotted in Fig.~\ref{02_reflectance}. Two clusters of objects are apparent in this plot, roughly associated with the S-type (high $a^*$ color) and C-type (low $a^*$ color) spectral categories. In Section \ref{high_astar}, we consider the cluster with $a^* > -0.015$, which is a recognizable dynamical family. The objects with $a^* < -0.015$ represent multiple dynamical families, as shown in Section \ref{low_astar}. This multiplicity is already indicated by the presence of a number of X-type objects in the region with $a^*$ colors intermediate between the S- and C-type ranges; in Fig.~\ref{02_reflectance} they appear as the subtle grouping between $-0.1 < a^* < -0.0$.

\begin {center}
EDITOR: PLACE FIGURE \ref{02_reflectance} HERE
\end {center}

\section{High-$a^*$ objects: The Hertha1 family}
\label{high_astar}

\subsection{Family identification and characterization}
\label{identify}

\begin{figure}
\vspace{-0.4in}
\begin{center}
\includegraphics [width=3.8in]{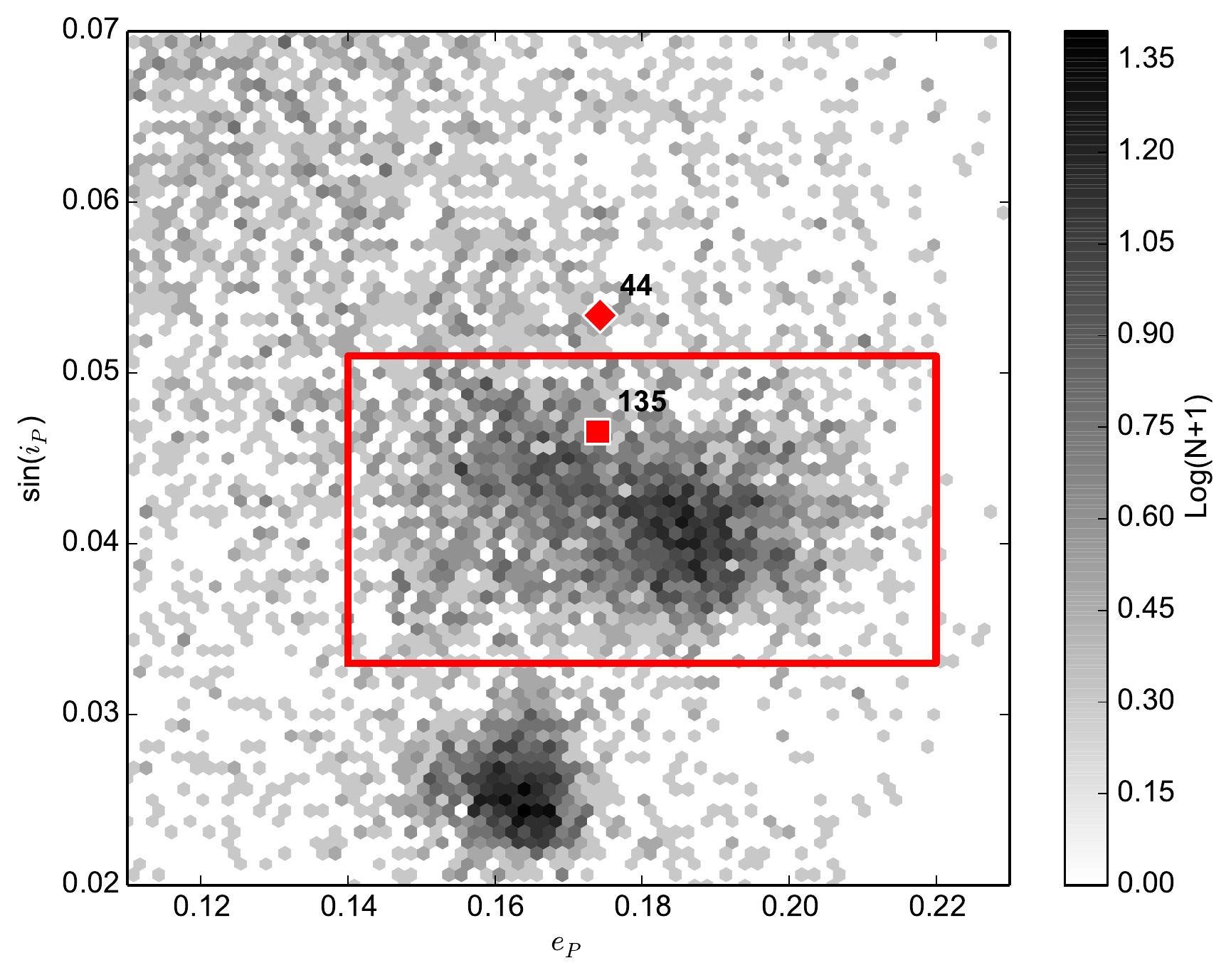}
\caption {\footnotesize Proper orbital elements for all objects in Fig.~\ref{02_reflectance} with $a^* > -0.015$. The large structure that occupies much of the box is here designated the Hertha1 family. As in Fig.~\ref{01_orbital}, the red diamond and square represent (44) Nysa and (135) Hertha, respectively. }
\label{03_high_astar}
\end{center}
\end{figure}

\begin{figure}
\begin{center}
\includegraphics [width=3.8in]{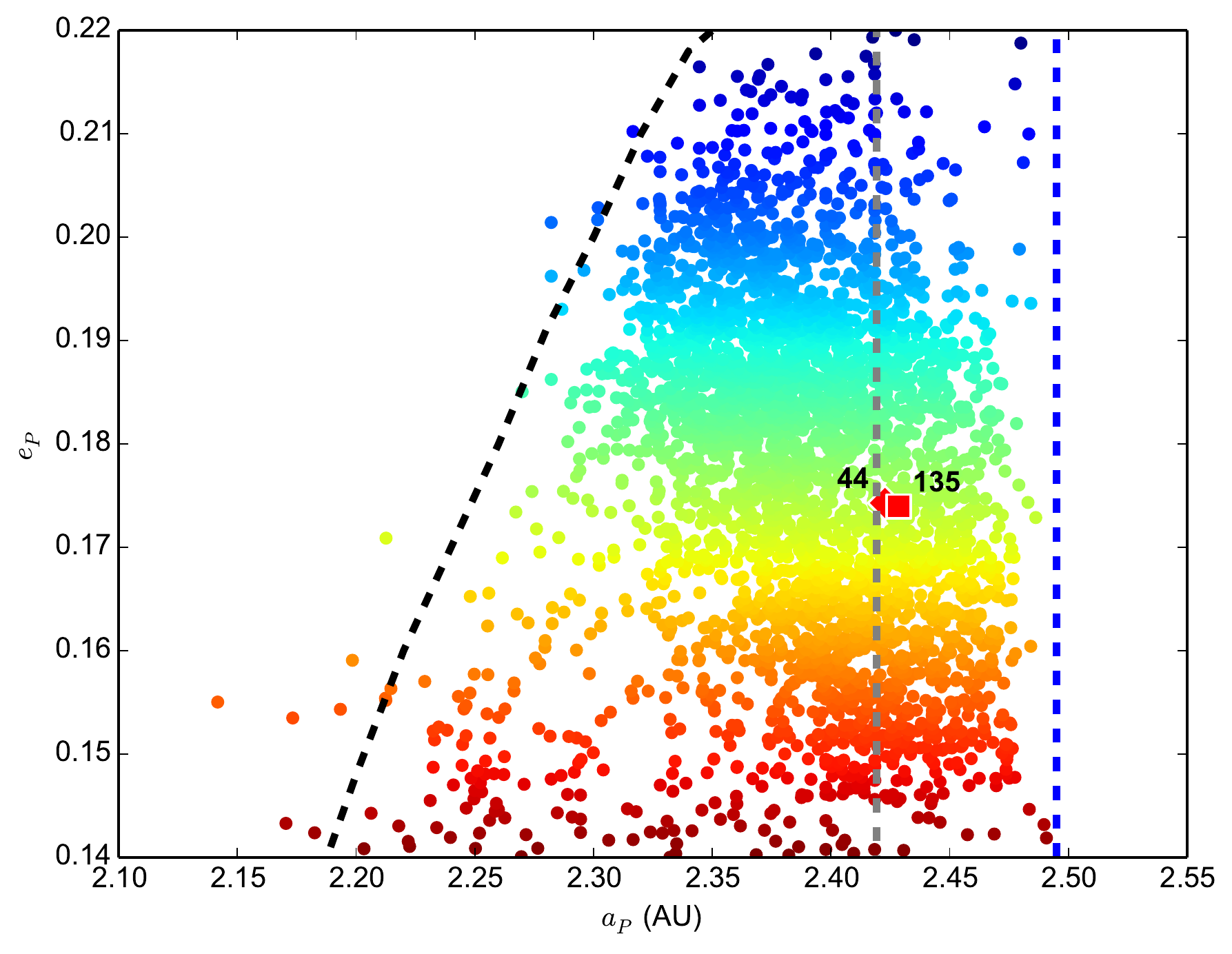}
\caption {\footnotesize Proper semimajor axis distribution for all of the objects within the red box in Fig.~\ref{03_high_astar}. The dotted black, blue and gray lines indicate the locations of the Mars-crossing line and the Jupiter 3:1 and Mars 1:2 mean-motion resonances, respectively. The Hertha1 objects show a trend of decreasing eccentricity for higher semimajor axes. At the lowest semimajor axes and highest eccentricities, the distribution has been truncated by the Mars-crossing line; at the highest semimajor axes and lowest eccentricities, the distribution has been truncated by the Jupiter 3:1. The most pristine ``boundary'' of the family occurs for the objects with 2.3~AU $< a_\text{P} <$ 2.4~AU and $e_\text{P} < 0.185$. The color gradient in eccentricity is used in Fig.~\ref{05_color}. }
\label{04_ae}
\end{center}
\end{figure}

The proper element distribution for the Nysa-Polana region objects from Fig.~\ref{02_reflectance} with $a^* > -0.015$ is shown in Figs.~\ref{03_high_astar} and \ref{04_ae}. These figures suggest that the high-$a^*$ structure (within the red box in Fig.~\ref{03_high_astar}) is associated dynamically with parent object (135) Hertha. We designate this cluster as the ``Hertha1'' family (cf. Table~\ref{table1} for previous designations in the literature), with the ``1'' added because (135) Hertha is also associated with a smaller cluster of objects (``Hertha2,'' described in Section \ref{low_astar}). 

\begin {center}
EDITOR: PLACE FIGURE \ref{03_high_astar} HERE
\end {center}

The distribution within the red box in Fig.~\ref{03_high_astar} could be interpreted as two separate clumps around $e_\text{P}$, $\sin i_\text{P}$ of (0.170, 0.045) and (0.185, 0.040). These two clumps were named the ``Hertha'' and ``Mildred'' families, respectively, by \citet{mothe-diniz2005}, cf. Table~\ref{table1}. However, the $a_\text{P}$-$e_\text{P}$ distribution of the objects within the red box (Fig.~\ref{04_ae}) shows a continuous and uniform trend of lower $e_\text{P}$ with increasing $a_\text{P}$, consistent with the ejection field from a single large-scale collision event. Moreover, as we show below, the Yarkovsky V shape is similarly continuous within and between the two populations. Thus we interpret the structure within the red box in Fig.~\ref{03_high_astar} as a single collisional family, Hertha1.

\begin {center}
EDITOR: PLACE FIGURE \ref{04_ae} HERE
\end {center}

\begin{figure}
\vspace{-0.5in}
\begin{center}
  \begin{subfigure}{\textwidth}
    \begin{centering}
    \includegraphics [width=3.8in]{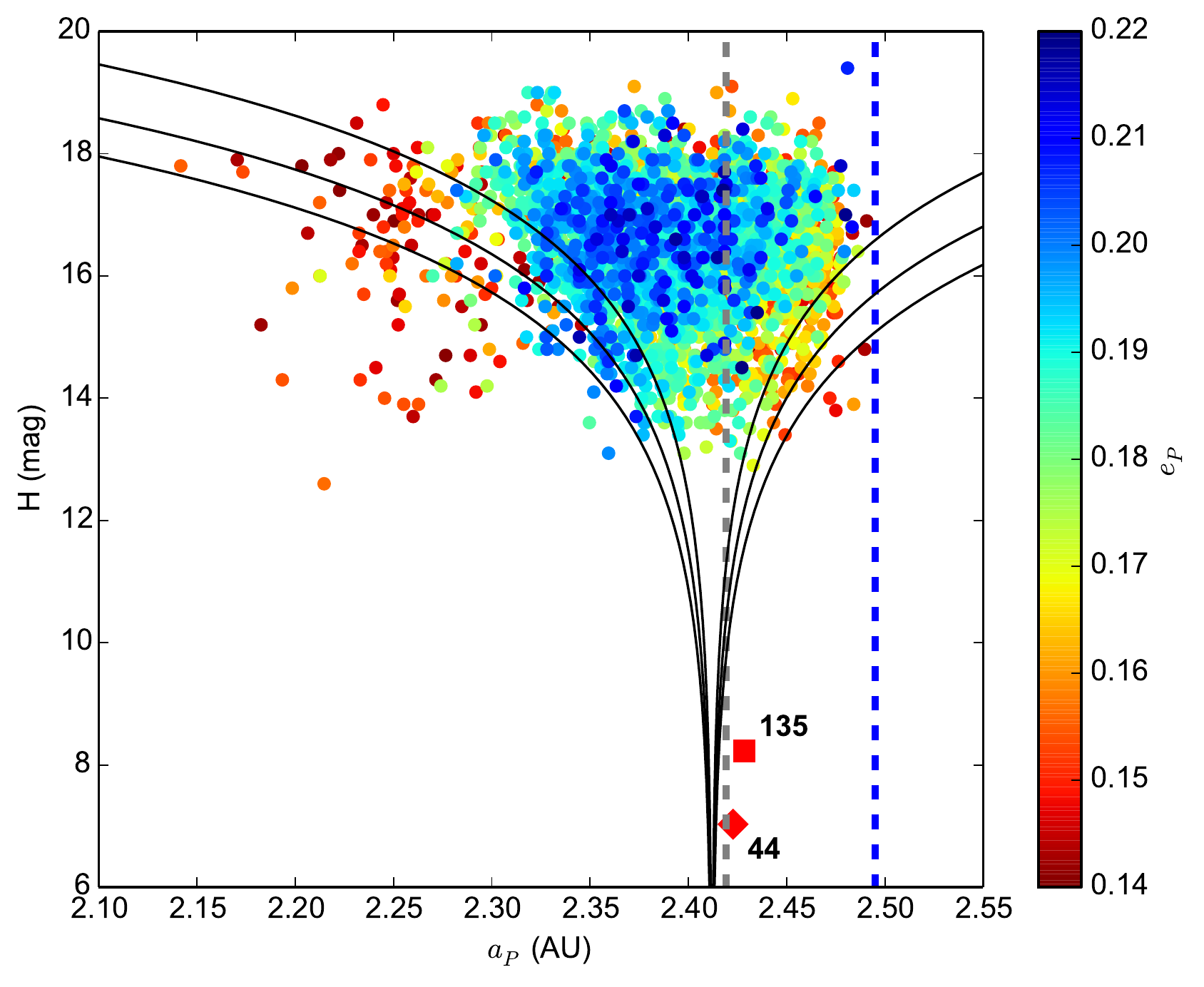}
    \vspace{-0.1in}
    \caption{}
    \label{05_color_a}
    \end{centering}
  \end{subfigure}
  \begin{subfigure}{\textwidth}
    \begin{centering}
    \includegraphics [width=3.8in]{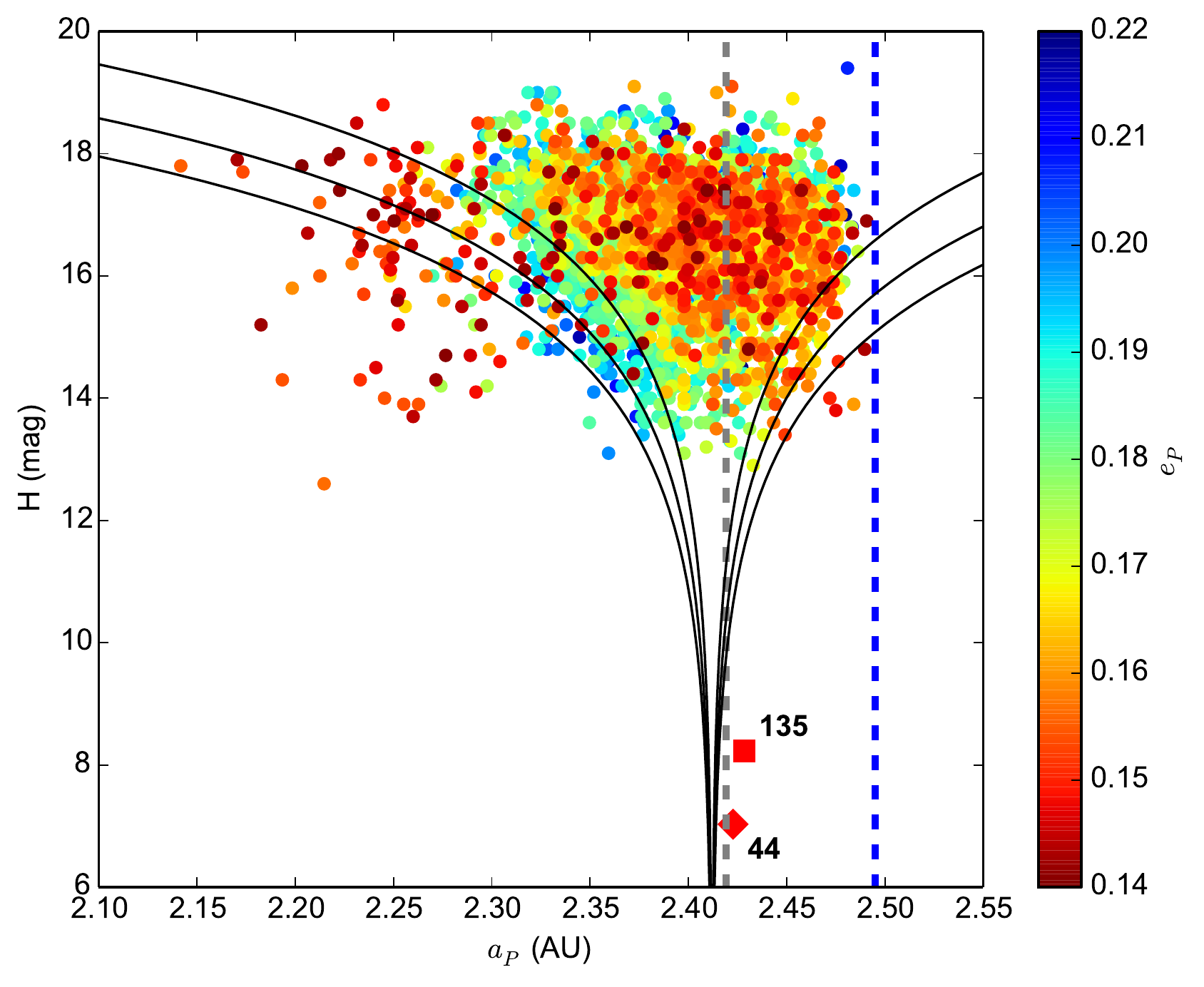}
    \vspace{-0.1in}
    \caption{}
    \label{05_color_b}
    \end{centering}
  \end{subfigure}
  \caption {\footnotesize The population from Fig.~\ref{04_ae}, plotted in $a_\text{P}-H$ space, with color indicating eccentricity. Plots (a) and (b) are identical except for the ordering of the objects plotted: In (a), higher $e_\text{P}$ objects are plotted on top of lower $e_\text{P}$ objects; in (b), the lower $e_\text{P}$ objects are plotted on top to show the behavior at the lower eccentricities. The Yarkovsky ``V'' shape is evident among the family members. The black lines represent $|C|$ = 0.04~mAU, 0.06~mAU and 0.08~mAU; none of the lines fit the outer edge of the distribution well, due to the $e_\text{P}$ dependence which skews the distribution in $a_\text{P}$.}
\label{05_color}
\end{center}
\end{figure}

The absolute magnitude distribution of the objects of the Hertha1 family yields information about its dynamical evolution under the influence of the Yarkovsky effect. The $a_\text{P}-H$ distribution plotted in Fig.~\ref{05_color} shows the V shape characteristic of a collisional family evolved under the Yarkovsky effect. The rightmost edge of the V shape is truncated by the Jupiter 3:1 mean motion resonance (represented by the blue dotted line), and the objects at high eccentricities and low semimajor axes are affected by encounters with Mars, modifying the distribution from an ``ideal'' V shape. The low-$a_\text{P}$, low-$e_\text{P}$ objects evident in Fig.~\ref{04_ae} are here shown to be primarily interlopers in the family, included in this sample only on the basis of their inclusion within the family ranges in $e_\text{P}$ and $\sin i_\text{P}$ and $a^*$.

\begin {center}
EDITOR: PLACE FIGURE \ref{05_color} HERE
\end {center}

In principle, the age of the family can be interpreted from the distribution of $C$ parameters of the family members (Equation \ref{Cparam}), the outer boundary of which describes the V-shaped envelope in $a_\text{P}$ vs. $H$. However, the correlation between $e_\text{P}$ and $a_\text{P}$ among the Hertha1 family objects (Fig.~\ref{04_ae}) means that the center of the Yarkovsky V shifts to lower $a_\text{P}$ for larger eccentricities. This complicates the identification of the Yarkovsky envelope that best fits the boundary of the family in semimajor axis; none of the example Yarkovsky envelopes plotted in Fig.~\ref{05_color} match the boundary well. The full semimajor axis width in Fig.~\ref{05_color} is due to the skewed $e_\text{P}$ vs. $a_\text{P}$ distribution in Fig.~\ref{04_ae}, which is related to the initial collision rather than the subsequent Yarkovsky drift. 

This skew must be removed prior to interpretation of the family's age via the observed Yarkovsky spread in $a_\text{P}$. To do this, we need to adjust each member's semimajor axis by an amount 

\begin{equation}
 a_\text{P}' - a_\text{P} = \frac{1}{m}\cdot (e_\text{P} - e_{\text{P,0}}),
 \label{aprime}
\end{equation}

\noindent where $m$ represents the slope of the $a_\text{P}$ vs. $e_\text{P}$ distribution, and $e_{\text{P,0}}$ is the family center in eccentricity. Given appropriate values for $m$ and $e_{\text{P,0}}$, we can apply a shear transformation to the distribution of objects to remove the eccentricity dependence and determine the Yarkovsky-induced spread in $a_\text{P}$.

In order to find the slope $m$ of the observed distribution, we must take into account the effects on the shape of the distribution due to the various resonances in and around the Nysa-Polana region (Fig.~\ref{04_ae}). The Mars-crossing region removes low-$a_\text{P}$, high-$e_\text{P}$ Hertha1 objects; the Jupiter 3:1 mean-motion resonance removes high-$a_\text{P}$, low-$e_\text{P}$ objects; lastly, the Mars 1:2 mean-motion resonance depletes family members both below $a_\text{P} = 2.43$~AU and above $e_\text{P} = 0.195$ as they drift outward in semimajor axes due to the Yarkovsky effect and exit the region before drifting back. 

\begin{figure}
\vspace{-0.4in}
\begin{center}
\includegraphics [width=3.8in]{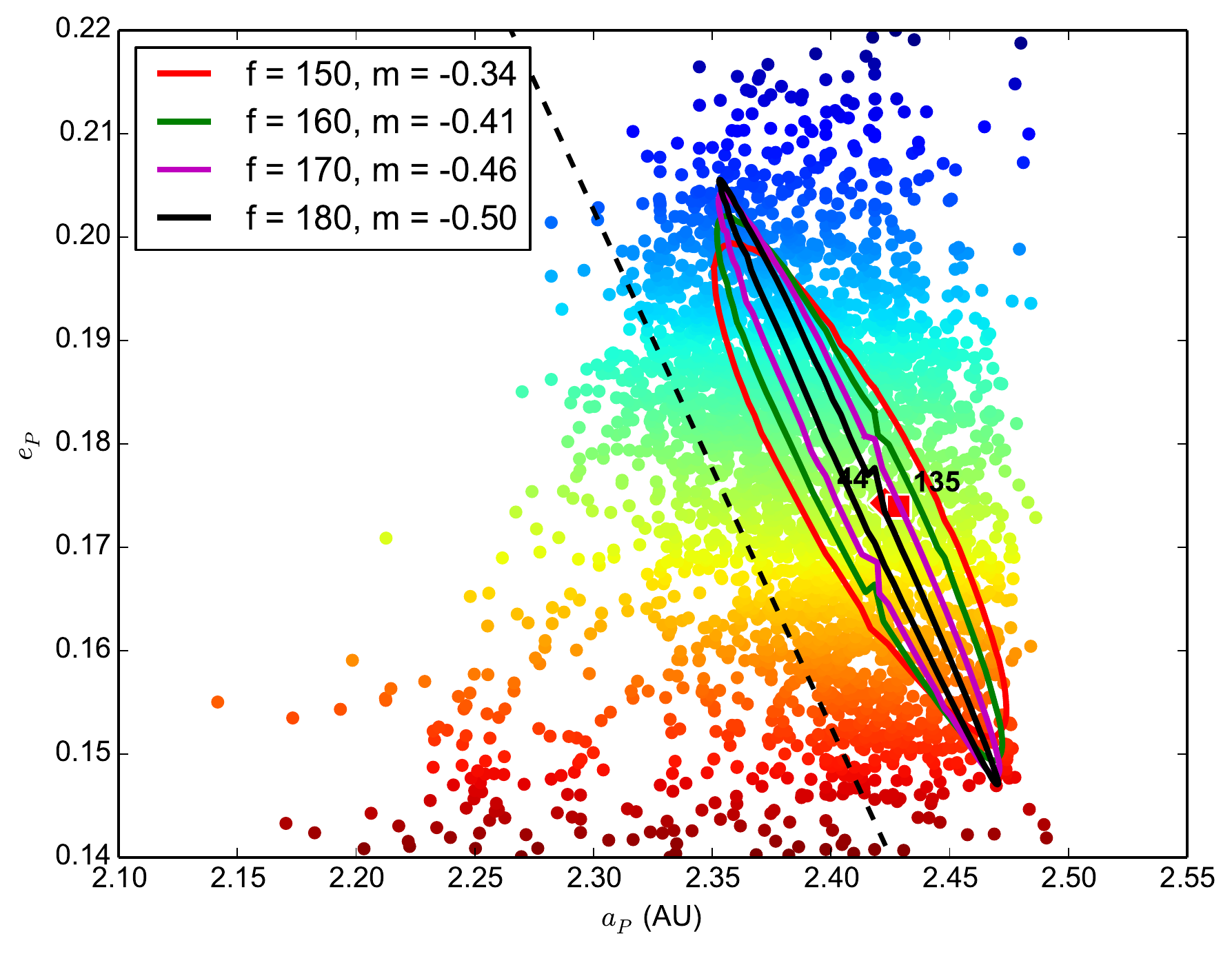}
\caption {\footnotesize Equivelocity ellipses (i.e. the distribution in $e_\text{P}, \sin i_\text{P}$ of particles ejected isotropically at 285~m/s from the barycenter of the population) for various values of the true anomaly, $f$, of the collision event, plotted over the distribution from Fig.~\ref{04_ae}. The best linear fit to the boundary, with slope $m = -0.50$~AU$^{-1}$, is also shown (dashed line on left).}
\label{07_gauss}
\end{center}
\end{figure}

The only ``pristine'' edge of the family that remains untouched by resonances (and thus still reflects the semimajor axis dependence of $e_\text{P}$ from the original collision) is the lower left edge, at $0.15 < e_\text{P} < 0.185$, where the sloped boundary of the family is visible (Fig.~\ref{04_ae}). We use this boundary to estimate the slope of the family; specifically, we find the boundary at each value of eccentricity within the range $0.15 < e_\text{P} < 0.185$, using a stepsize of $\delta e = 0.001$ and an eccentricity width of $\Delta e = 0.01$. The latter parameter specifies the window in eccentricity which defines the sample of objects used to find the boundary at each eccentricity step. The boundary is defined for this purpose as the point of maximum increase in density of objects, or a maximum in the derivative of the density function (see Appendix A.1 of \citet{dykhuis2014} for discussion of the kernel density estimation and its derivatives). A linear fit to the boundary values yields a slope of $m = (-0.50 \pm 0.04)$~AU$^{-1}$, represented by the dotted line in Fig.~\ref{07_gauss}.

The slope $m$ is a signature of the semimajor axis and eccentricity dispersion of the Hertha1 family that resulted from the original collision. In particular, $m$ is determined by the true anomaly, $f$, of the parent asteroid at the time of its disruption. This relationship is described by Gauss' equations, which give the change in orbital elements for a collision fragment that experiences a velocity impulse $\Delta v$ \citep[see, e.g.][]{zappala2002}. Gauss' equations describe the behavior of the collision fragments in unperturbed (osculating) elements; in the Hertha region, the proper elements show similar behavior. 

We plot the proper element distributions for several different values of $f$ in Fig.~\ref{07_gauss} \citep[cf.][Fig.~1]{nesvorny2002}, along with their corresponding slopes, $m$. Each ellipse represents an isotropic ejection field with $\Delta v = 285$~m/s, with $\Delta v$ chosen to approximate the eccentricity range of the family, estimated by the noticeable decrease in number density near $e_\text{P}$ = 0.15 and $e_\text{P}$ = 0.205. The ellipses were generated from 100 test particles, with osculating orbits calculated via the Gauss equations, and proper elements calculated via two-million-year integrations using Orbit9. The effects of the Mars 1:2 mean-motion resonance at $a_\text{P} = 2.419$~AU are observed in the proper element calculations. The lowest possible value for $m$ occurs at $f = 180\degree$, where $m = -0.50$~AU$^{-1}$. Thus, assuming the original ejection velocity field was fairly isotropic, it appears the collision that formed this family occurred when the parent body was near aphelion, with a true anomaly within the range $170\degree < f < 190\degree$, calculated using the uncertainty on the slope as determined from the boundary values above.

\begin {center}
EDITOR: PLACE FIGURE \ref{07_gauss} HERE
\end {center}

In order to apply the correction (Eq. \ref{aprime}) to remove the effect of the initial slope in the ($e_\text{P}$,$a_\text{P}$) distribution, we need to evaluate $e_{\text{P,0}}$ in addition to $m$. The value $e_{\text{P,0}}$ is the proper eccentricity at the family's center, which we take to be the barycenter of the population. To minimize the influences of the nearby resonances on the barycenter, we consider only the largest objects (12.0~mag $< H <$ 14.0~mag), which have experienced the least amount of Yarkovsky semimajor axis dispersion toward significant resonances. We further restrict the sample to 2.31~AU $< a_\text{P} <$ 2.5~AU, 0.136 $< e_\text{P} <$ 0.23, and 0.033 $< \sin i_\text{P} <$ 0.051, in order to conservatively include all Hertha1 family members while largely eliminating members of the nearby Massalia and Flora families. We also removed from the sample all objects with albedos lower than 0.14, which are likely to be associated with the overlapping darker families discussed in Section \ref{low_astar}. This does not remove all of the contamination from those families, due to the incompleteness of the WISE database; however, such contamination would likely be primarily from smaller objects, and thus would have little effect on the determination of the barycenter of the Hertha1 family. Lastly, we removed asteroid (135) Hertha itself from the barycenter consideration, due to the uncertainty surrounding its inclusion in the family. 

\begin{figure}
\begin{center}
\includegraphics [width=3.8in]{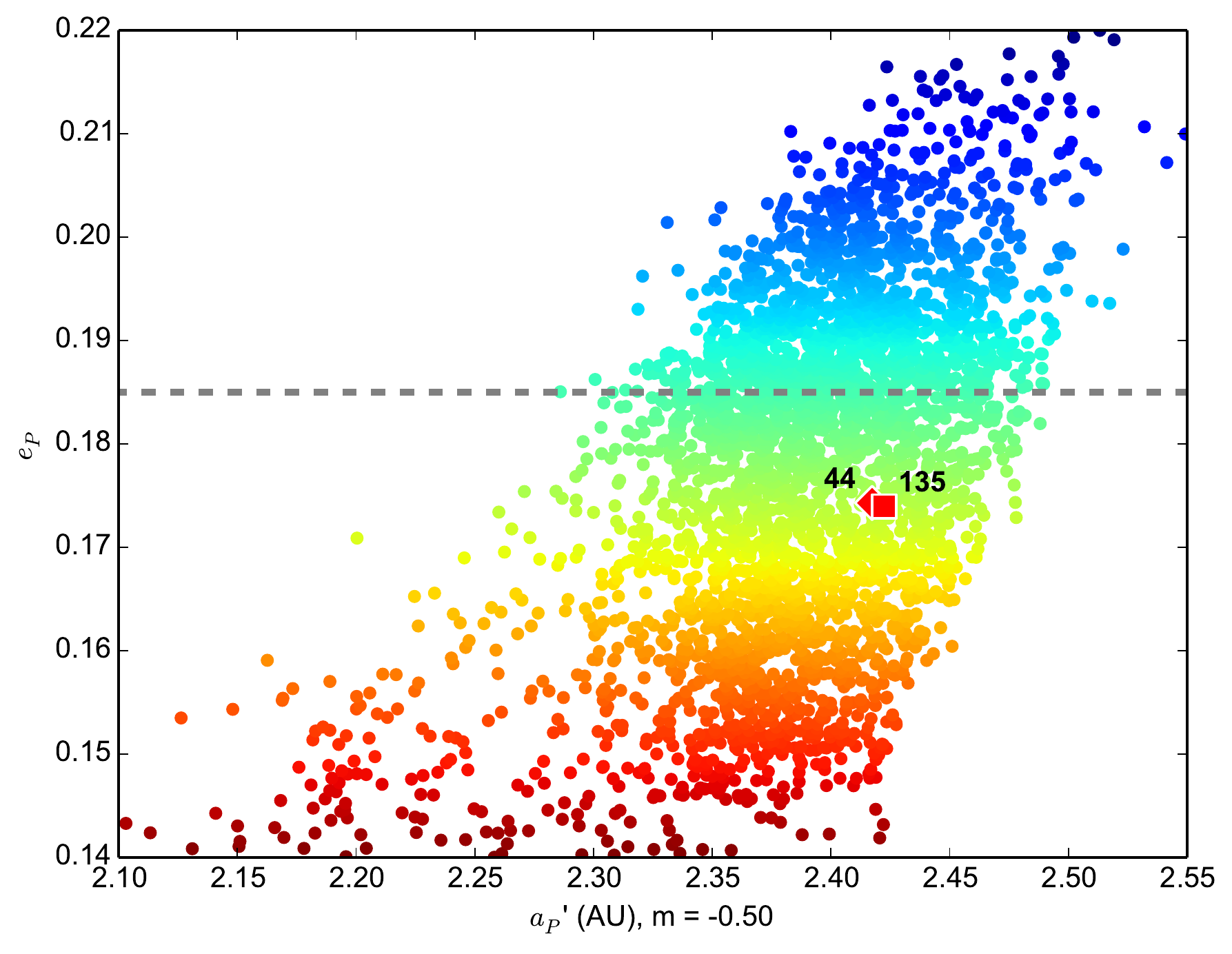}
\caption {\footnotesize The $e_\text{P}, a_\text{P}$ distribution of the objects of the Hertha1 family, with the corrected values for proper semimajor axis ($a_\text{P}'$, calculated from Eq. \ref{aprime} using slope $m = -0.50$~AU$^{-1}$ and $e_\text{P,0} = 0.177$), removing the skew in the distribution due to the initial collision. The boundary along the left edge is now vertical. The Mars-crossing region and Jupiter 3:1 resonance have noticeably depleted the objects at high $e_\text{P}$ (blue dots) and high $a_\text{P}$ (right side of the family), respectively. }
\label{08_aprime_e}
\end{center}
\end{figure}

The barycenter of the sample described above is $a_{\text{P,0}} = 2.412 \pm 0.015$~AU and $e_{\text{P,0}} = 0.177 \pm 0.003$, with the uncertainty dominated by the exclusion of (135) Hertha from the sample. The corrected $e_\text{P}$ vs. $a_\text{P}'$ distribution, with $a_\text{P}'$ calculated using Eq. \ref{aprime} with $m = -0.50$~AU$^{-1}$ and $e_{\text{P,0}} = 0.177$, is shown in Fig.~\ref{08_aprime_e}.

\begin {center}
EDITOR: PLACE FIGURE \ref{08_aprime_e} HERE
\end {center}

\subsection{Post-collision history of the family}
\label{interpret}

The corrected semimajor axis dispersion of the Hertha1 family records information about the family's post-collision evolution under the Yarkovsky effect. In general, the most useful Yarkovsky drift information is preserved in the outer edges of a family, specifically the extent of drift at each size range, as parameterized by $C$. The age of the family is obtained from the value of $C$ for which Eq. \ref{Cparam} gives curves in $a_\text{P}'-H$ space that best fit the outer edges of the distribution. However, the Jupiter 3:1 resonance at high $a_\text{P}$ and the Mars-crossing region at low $a_\text{P}$ have significantly sculpted the boundaries of the Hertha1 family (Fig.~\ref{08_aprime_e}). The former renders the high-$a_\text{P}$ (right) edge of the family unreliable; the Mars-crossing region affects the low-$a_\text{P}$ (left) edge for $e_\text{P} > 0.185$. Accordingly, in order to determine an appropriate boundary $C$ value, we consider only the objects with $e_\text{P} < 0.185$, and fit only the lower-$a_\text{P}$ edge of their distribution in $a_\text{P}'-H$ space.

\begin{figure}
\begin{center}
\includegraphics [width=3.8in]{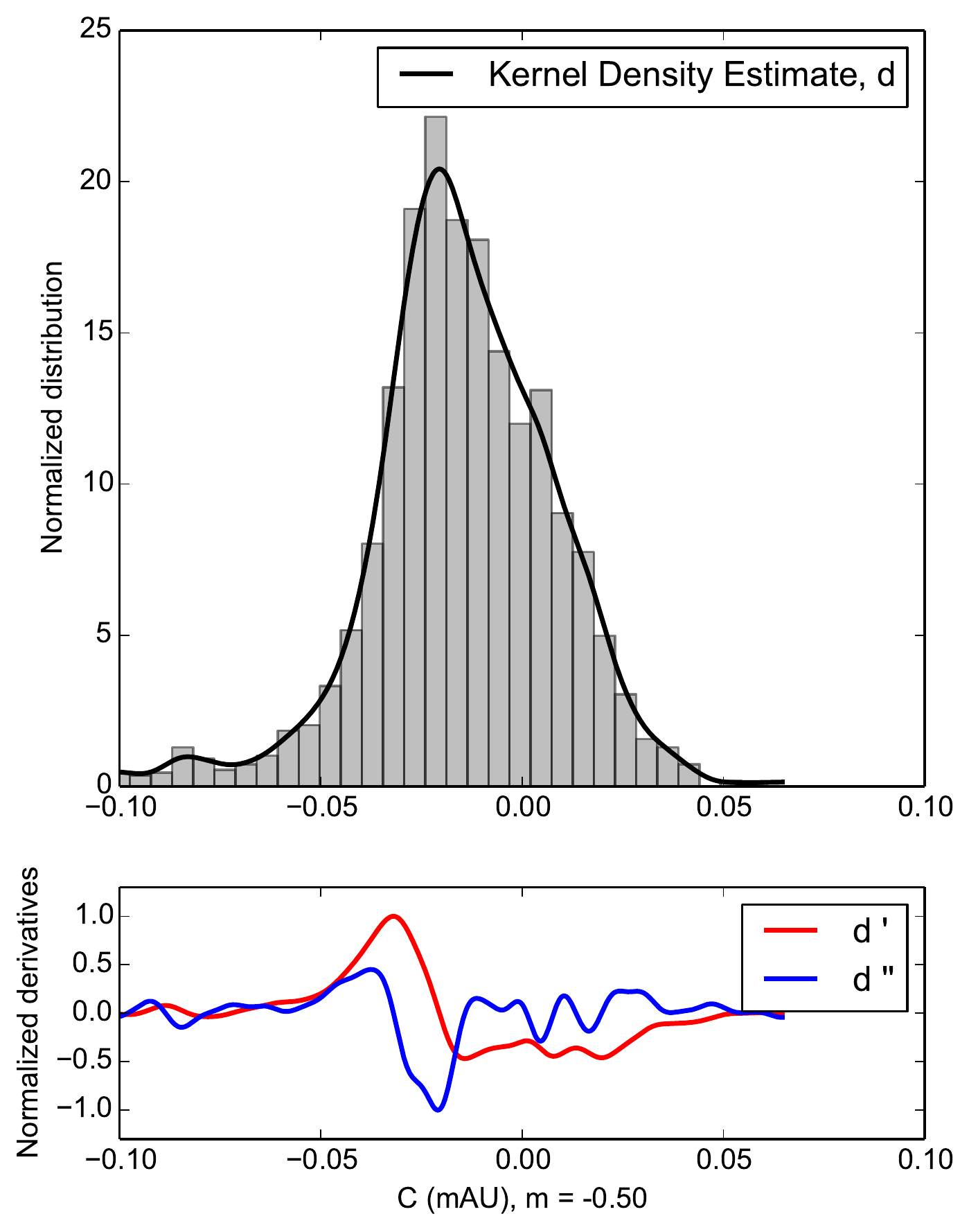}
\caption {\footnotesize The $C$ distribution of the Hertha1 family, computed from Eq. \ref{Cparam}, using the corrected $a_\text{P}'$ with slope $m = -0.50$~AU$^{-1}$. The curves represent the kernel density estimate (KDE) for the distribution, along with its first and second derivatives. The best-fit lower $C$ boundary occurs at a local maximum in the second derivative, $C = -0.045$~mAU. }
\label{09_cprime}
\end{center}
\end{figure}

To determine the value of $C$ that best fits that edge, we calculate for each asteroid the value of $C$ that corresponds to its $a_\text{P}'$ and $H$ values (Eq. \ref{Cparam}). Fig.~\ref{09_cprime} shows the distribution of these $C$ values, with the curves representing the kernel density estimate (KDE) for the distribution along with its derivatives (see Appendix A.1 of \citet{dykhuis2014} for an explanation of the KDE and its derivatives). The best-fit edge of the family is the $C$ value where there is a maximum in the second derivative of the KDE (the blue curve in Fig.~\ref{09_cprime}). 

\begin {center}
EDITOR: PLACE FIGURE \ref{09_cprime} HERE
\end {center}

\begin{figure}
\vspace{-0.7in}
\begin{center}
  \begin{subfigure}{\textwidth}
    \begin{centering}
    \includegraphics [width=3.8in]{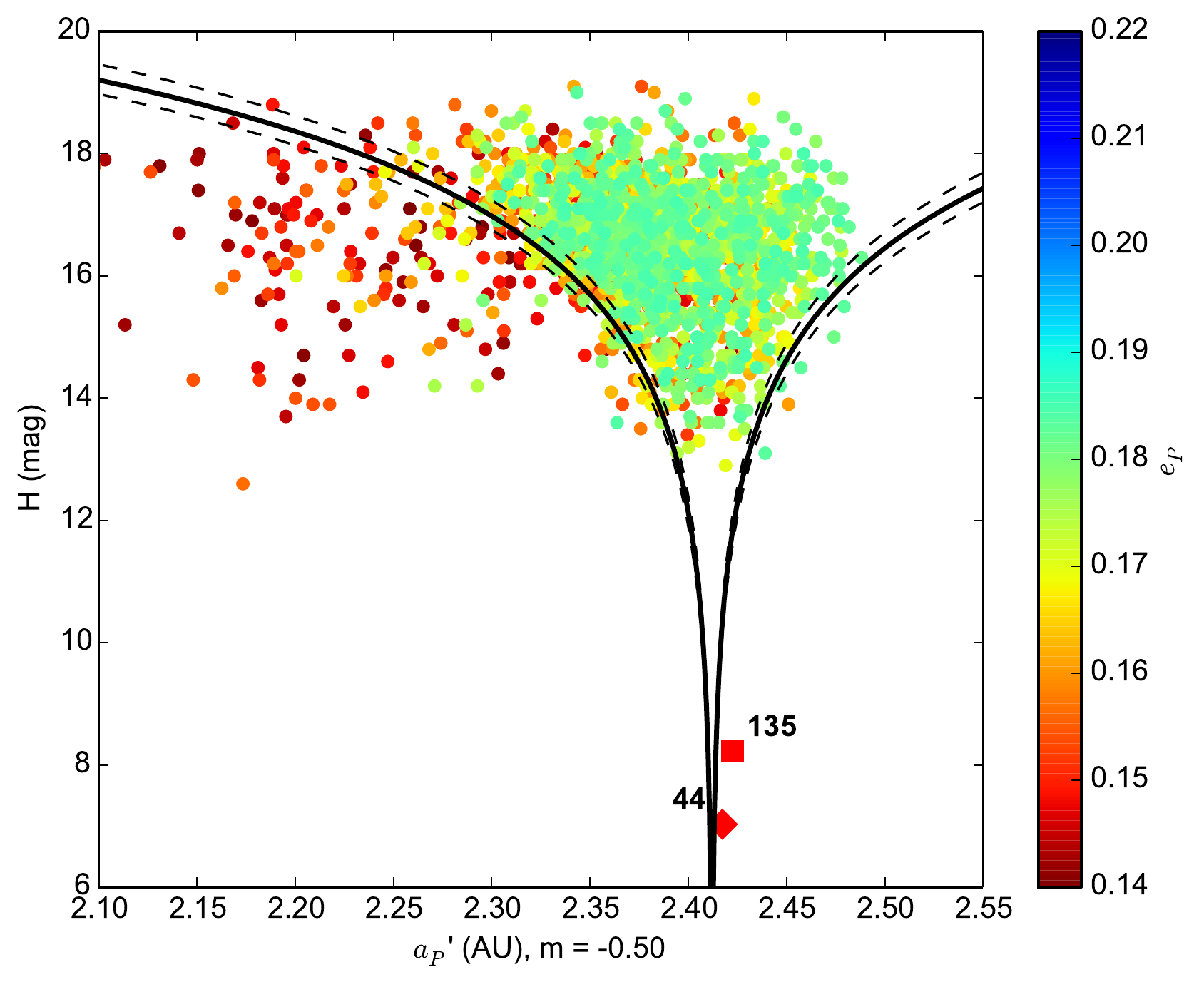}
    \vspace{-0.1in}
    \caption{}
    \label{10_aprime_H_a}
    \end{centering}
  \end{subfigure}
  \begin{subfigure}{\textwidth}
    \begin{centering}
    \includegraphics [width=3.8in]{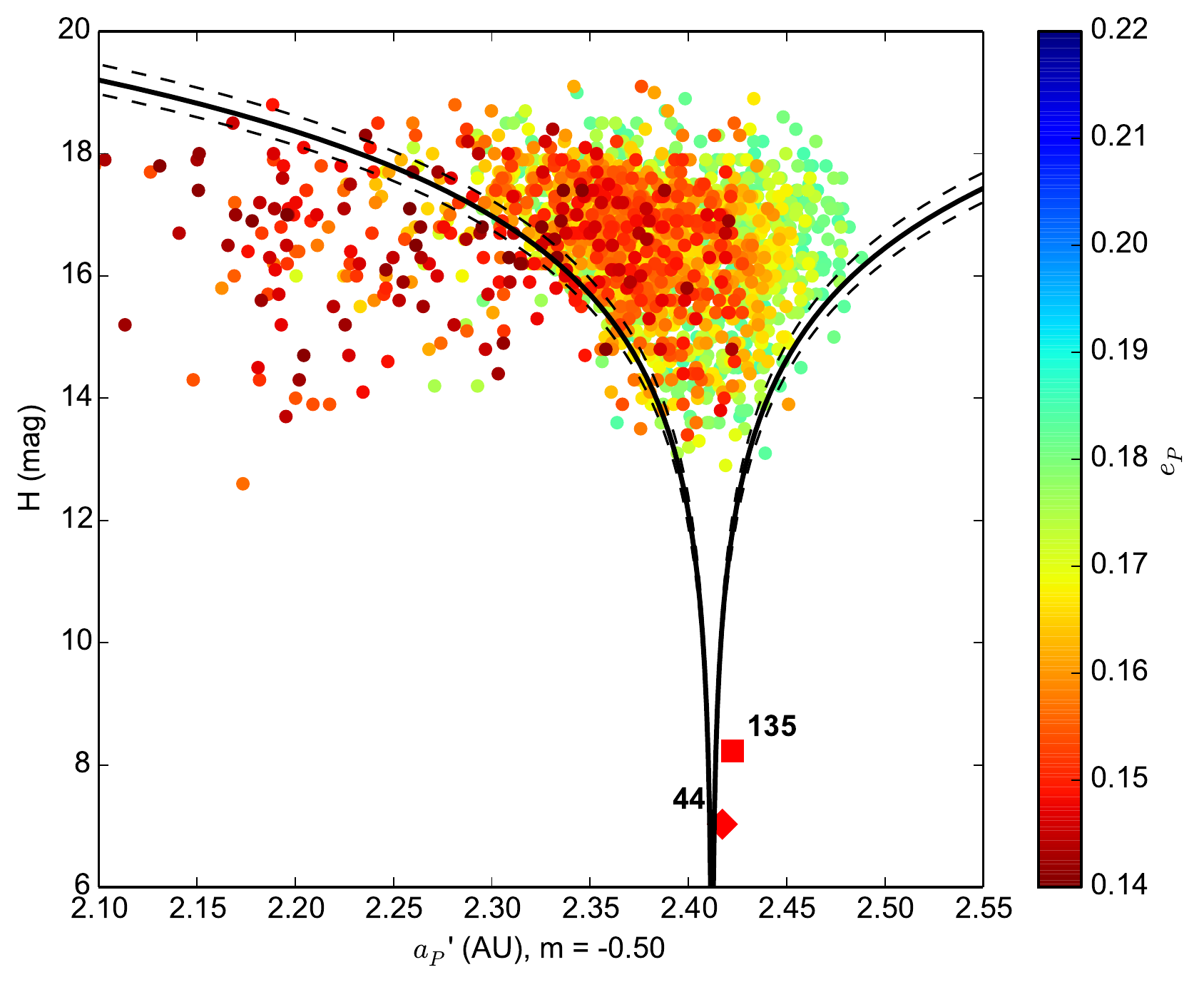}
    \vspace{-0.1in}
    \caption{}
    \label{10_aprime_H_b}
    \end{centering}
  \end{subfigure}
  \caption {\footnotesize The distribution of Hertha1 objects in $a_\text{P}'-H$ space. Objects with $e_\text{P} > 0.185$ are not included to avoid the effect of the Mars-crossing region on the left edge boundary. The black lines shows the amount of semimajor axis drift that correspond to $C(a_\text{P}')$ values of $\pm0.045$~mAU, with dotted lines representing the uncertainty of $\pm$ 0.005~mAU. The curves are fit to the left edge of the population; the curves on the right side are mirror images of those on the left for positive $C$ values. }
\label{10_aprime_H}
\end{center}
\end{figure}

There are comparable maxima in the second derivative at $C = -0.045$~mAU and $-0.037$~mAU; however, inspection of the distribution in $a_\text{P}'-H$ space (Fig.~\ref{10_aprime_H}) shows that the latter boundary would lie well inside the dense population, not at its edge. By contrast, $C = -0.045$~mAU (black curve in Fig.~\ref{10_aprime_H}) does indeed fit the outer edge of the distribution quite well. The uncertainty in the $C$ boundary ($\pm 0.005$~mAU, represented by the dashed curves in Fig.~\ref{10_aprime_H}) is dominated by the uncertainty in the $a_{\text{P,0}}$ center of the family. The $C$ boundary defining the edge of the family is much tighter for the corrected $a_\text{P}'-H$ distribution than it was for the uncorrected $a_\text{P}-H$ distribution plotted earlier in Fig.~\ref{05_color} ($C = \pm0.045$~mAU vs. $C = \pm0.075$~mAU).

\begin {center}
EDITOR: PLACE FIGURE \ref{10_aprime_H} HERE
\end {center}

The value of $C = 0.045$~mAU fits the leftmost edge of the distribution quite well for the larger objects ($H <$ 17~mag), but fails for the smallest objects ($H >$ 17 mag), which is not surprising for a couple of reasons. First, the observational bias in the database affects the completeness of the family sample for objects fainter than magnitude 16.5 \citep{bottke2014}. Second, the effects of ``variable'' or ``stochastic'' YORP on the drifting family members can cause their spin rates to random walk toward the typical YORP end states (mass shedding or non-principal axis rotation), which applies a size-dependent ``brake'' to the smallest objects \citep{bottke2013,bottke2014}. Stochastic YORP simulations of the Eulalia, ``New Polana'' and Erigone families by these authors demonstrated behavior similar to what is observed here. 

The calculation of the age of the family from the observed Yarkovsky semimajor axis dispersion requires an estimate of the initial semimajor axis spread due to the original collision itself. This estimate is typically difficult to obtain, but in the case of the Hertha1 family, we can place an upper limit on the collision spread from the $a_\text{P}-e_\text{P}$ distribution. As shown in Section \ref{identify}, the slope of $m = -0.50$~AU$^{-1}$ indicates a collision near aphelion, which would result in minimal initial semimajor axis dispersion in $a_\text{P}'$. The maximum expected value of dispersion is $\delta a_{ej} = 0.005 \pm 0.008$~AU, which represents the width of the proper semimajor axis dispersion for a collision at true anomaly $f = 180\degree$, with uncertainty derived from the uncertainty in $f$ (which results from the uncertainty in the slope $m$).

The semimajor axis dispersion due to the original collision is thus expected to be a small fraction of the Yarkovsky dispersion, resulting in an overestimation of the boundary $C$ parameter of less than 0.003 mAU, which is less than the uncertainty on the boundary $C$ parameter (0.005 mAU). Hence the initial semimajor axis dispersion has little effect on our calculation of the age from the Yarkovsky drift.

We use the inner boundary of the $C$ parameter distribution to determine the age of the family via the relation \citep{dykhuis2014}:

\begin{equation}
\label{yarcora3}
t = \frac{1329 \cdot |C| \cdot {a_{\text{P,0}}}^2}{c_Y \sqrt{p_\text{V}}(1-p_\text{V})\cos\epsilon}
\end{equation}

\noindent where $t$ is the time (in My) since the collision, $a_\text{P,0}$ is the time-averaged proper semimajor axis of the family members (in AU), $\epsilon$ is the obliquity, and $C$ is given in AU. The parameter $c_Y$ contains information about the asteroid's material properties (thermal conductivity, specific heat, material density), which are not known \emph{a priori}. As in \citet{dykhuis2014}, we adopt the value for $c_Y$ obtained from measurements of the precession rates of the Karin family, which yield a $c_Y$ parameter in the range of 0.0026 AU$^3$km~My$^{-1} < c_Y <$ 0.0035~AU$^3$km~My$^{-1}$. 

With values of $C = \pm0.045$~mAU, albedo $p_\text{V} = 0.25$, $a_{\text{P,0}} = 2.412$~AU, and $c_Y$ = 0.00305~AU$^3$km~My$^{-1}$, we find an age for the Hertha1 family of $300^{+60}_{-50}$~My. The error in this estimate is dominated by the uncertainty in $c_Y$. 

The age estimate also does not account for possible systematic errors due to the assumption of similar material properties between the members of the Hertha1 and Karin families. In particular, differences in bulk densities would affect the $c_Y$ parameter appropriate to both families. The average bulk densities for the two families are difficult to estimate, due to the lack of available density data for the family members. The only Hertha family object with density data is (135) Hertha itself \citep{carry2012}; its taxonomic type of X$_k$ makes it a poor standard for comparison with the S-type Karin family. The Karin family itself has members with density measurements, but the Koronis family, of which Karin is a subfamily, has density measurements for two of its members, (243) Ida and (720) Bohlinia; their densities are 2.35 $\pm$ 0.29 and 2.74 $\pm$ 0.56 g/$\text{cm}^3$, respectively \citep{carry2012}. These are both consistent with the averages for the S-type class, and thus are assumed here to be consistent with the unknown Hertha family average.

\section{Low-$a^*$ objects: The Hertha2, Polana and Eulalia families}
\label{low_astar}

\begin{figure}
\begin{center}
\includegraphics [width=3.8in]{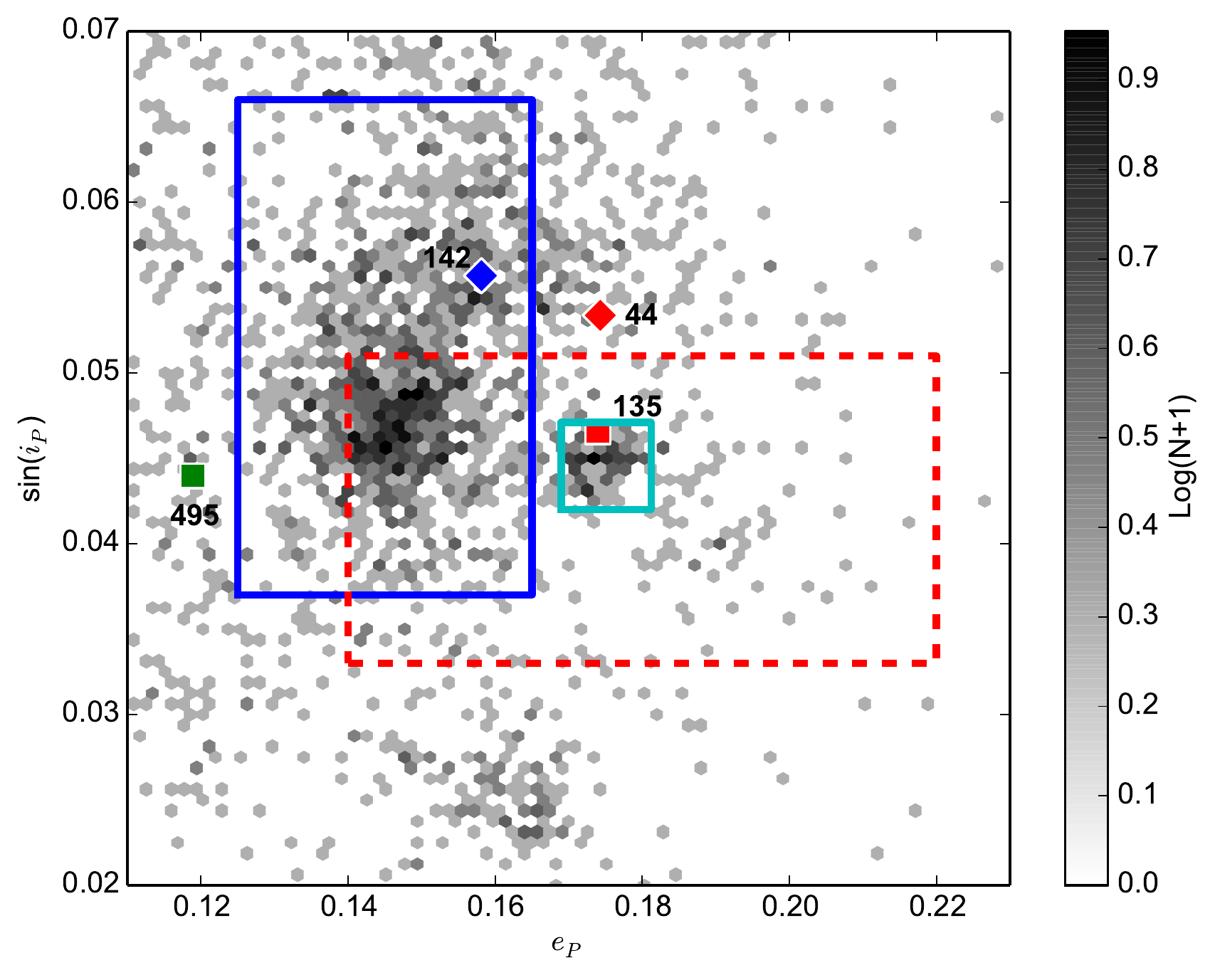}
\caption {\footnotesize Proper orbital elements for all objects in Fig.~\ref{02_reflectance} with $a^* < -0.015$. Two structures appear within the blue box: the diffuse Polana family surrounding (142) Polana and the larger, denser Eulalia1 family near $e_\text{P} = 0.145$ and $\sin i_\text{P} = 0.047$. Asteroid (495) Eulalia itself (green square) is offset in eccentricity from its nominal family. A cluster of low-$a^*$ objects (cyan box) is revealed near (135) Hertha, and is here designated ``Hertha2.'' The high-$a^*$ Hertha1 family totally obscures this cluster in Fig.~\ref{01_orbital}. The faint structure at about $\sin i = 0.025$ is a low-$a^*$ remnant of the Massalia family, the spread of whose distribution in $a^*$ overlaps the $a^* = -0.015$ boundary. }
\label{11_low_astar}
\end{center}
\end{figure}

We now consider the objects in the Nysa-Polana region with $a^* < -0.015$. The ($e_\text{P}$, $\sin i_\text{P}$) distribution of these objects is plotted in Fig.~\ref{11_low_astar}. Several structures appear within the blue box; the diffuse grouping near (142) Polana and the dense grouping near $e_\text{P} = 0.145$ and $\sin i_\text{P} = 0.047$ are here referred to as the Polana and Eulalia1 families, respectively (Table~\ref{table1}). The combined families had long been recognized as a single ``Polana'' family until their distinction in the work of \citet{walsh2013}.

\begin {center}
EDITOR: PLACE FIGURE \ref{11_low_astar} HERE
\end {center}

Comparison of Fig.~\ref{11_low_astar} with Figs.~\ref{01_orbital} and \ref{03_high_astar} shows that the Hertha1 family, roughly bounded by the red box, has been removed. However, a smaller, dense cluster of low-$a^*$ objects lies near the location of asteroid (135) Hertha, within the range of the now-removed Hertha1 family. We refer to this cluster as Hertha2.

\subsection{The Hertha2 family}
\label{hertha2}

\begin{figure}
\vspace{-0.2in}
\begin{center}
\includegraphics [width=3.8in]{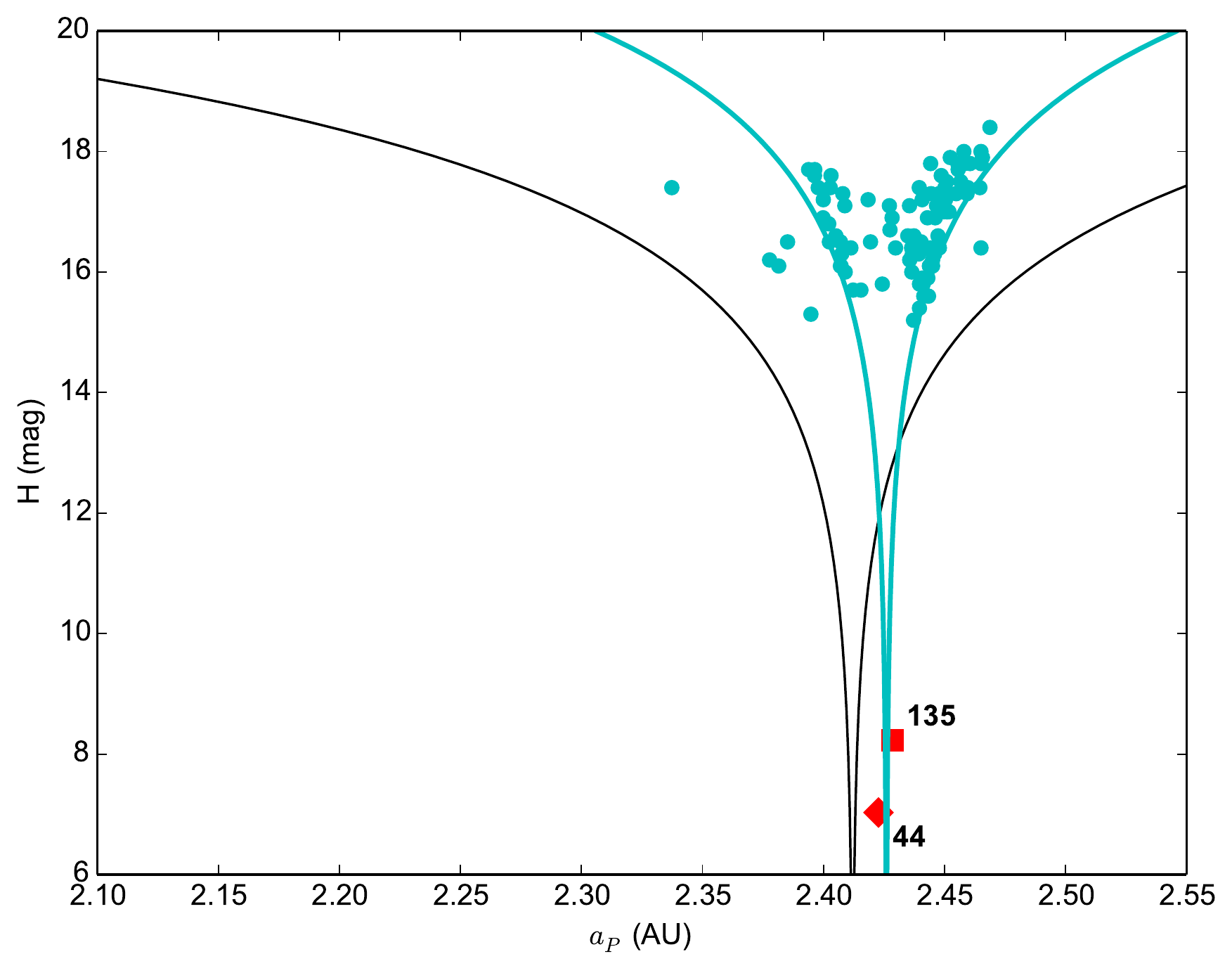}
\caption {\footnotesize The Hertha2 objects within the cyan box in Fig.~\ref{11_low_astar} are plotted in $a_\text{P}-H$ space. They display a Yarkovsky ``V'' bounded by $C = \pm0.012$~mAU (cyan lines). The $C$ parameter boundaries for the Hertha1 family from Fig.~\ref{10_aprime_H} are plotted as black lines for reference. Asteroid (135) Hertha is a plausible parent of the Hertha2 family as well as of Hertha1, although both families have distinct $C$ envelopes as well as reflectance properties (Fig.~\ref{13_hertha_region}). }
\label{12_hertha2_aH}
\end{center}
\end{figure}

The Hertha2 cluster lies very close to asteroid (135) Hertha in $a_\text{P}$ (Fig.~\ref{12_hertha2_aH}), as well as in $e_\text{P}$ and $\sin i_\text{P}$, suggesting a genetic relationship. In fact, the barycenter of the objects within the cyan box in Fig.~\ref{11_low_astar} (those plotted in Fig.~\ref{12_hertha2_aH}) is at $a_\text{P} = 2.426$~AU, nearly equal to the value for (135) Hertha itself, $a_\text{P} = 2.429$~AU. The cluster displays the characteristic V-shape in $a_\text{P}-H$ indicative of Yarkovsky evolution, further identifying the Hertha2 structure as a collisional family. The V-shape envelope corresponds to $C = \pm0.012$~mAU (significantly narrower than that observed for the Hertha1 family, $C$ = 0.045~mAU, Section \ref{high_astar}). This tighter V could represent either a younger age or a higher average density for the Hertha2 family; the implications are discussed further in Section \ref{discussion}.

\begin {center}
EDITOR: PLACE FIGURE \ref{12_hertha2_aH} HERE
\end {center}

\begin{figure}
\vspace{-0.2in}
\begin{center}
\includegraphics [width=3.8in]{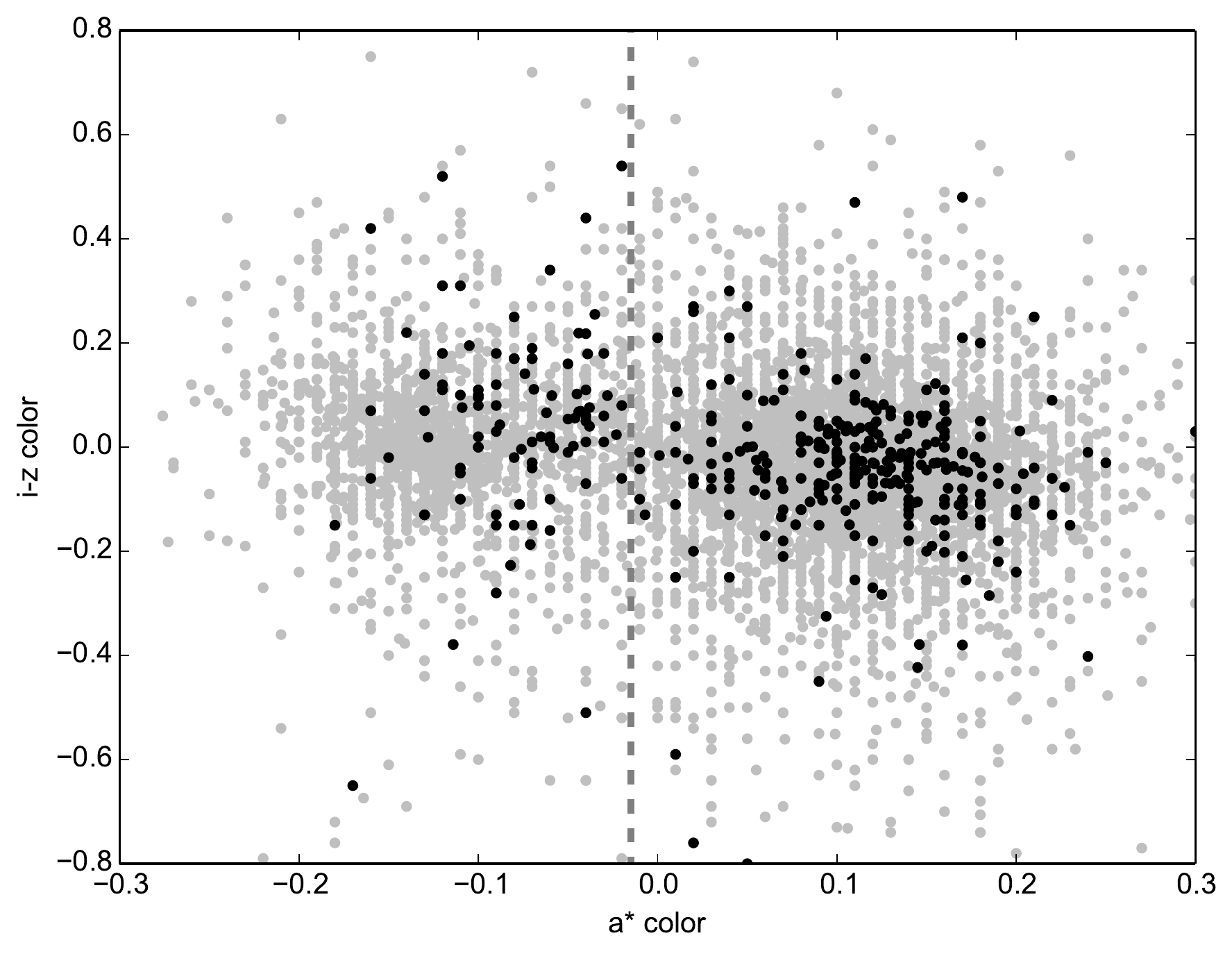}
\caption {\footnotesize The same as Fig.~\ref{02_reflectance}, but restricted to the range of $e_\text{P}$ and $\sin i$ given by the cyan box in Fig.~\ref{11_low_astar}. The Hertha2 family (near $a^* = -0.07$) is distinct from both the larger Hertha1 family (near $a^* = 0.13$) and the Polana/Eulalia families (near $a^* = -0.13$).}
\label{13_hertha_region}
\end{center}
\end{figure}

The Hertha2 family also appears as a distinct cluster in reflectance properties. Fig.~\ref{13_hertha_region} shows the color distribution similar to Fig.~\ref{02_reflectance}, but restricted to the objects within the same dynamical range as the Hertha2 family (2.1~AU $< a_\text{P} <$ 2.5~AU, $0.1690 < e_\text{P} < 0.1812$, and $0.0420 < \sin i_\text{P} < 0.0471$). The Hertha2 family forms a diffuse grouping centered around $a^* = -0.07$; the members of the large Hertha1 family that reside in the same dynamical space are distinct from the Hertha2 members, clustering instead near $a^* = 0.13$. 

\begin {center}
EDITOR: PLACE FIGURE \ref{13_hertha_region} HERE
\end {center}

\subsection{Other dark families}
\label{other_dark}

\begin{figure}
\vspace{-0.3in}
\begin{center}
\includegraphics [width=3.8in]{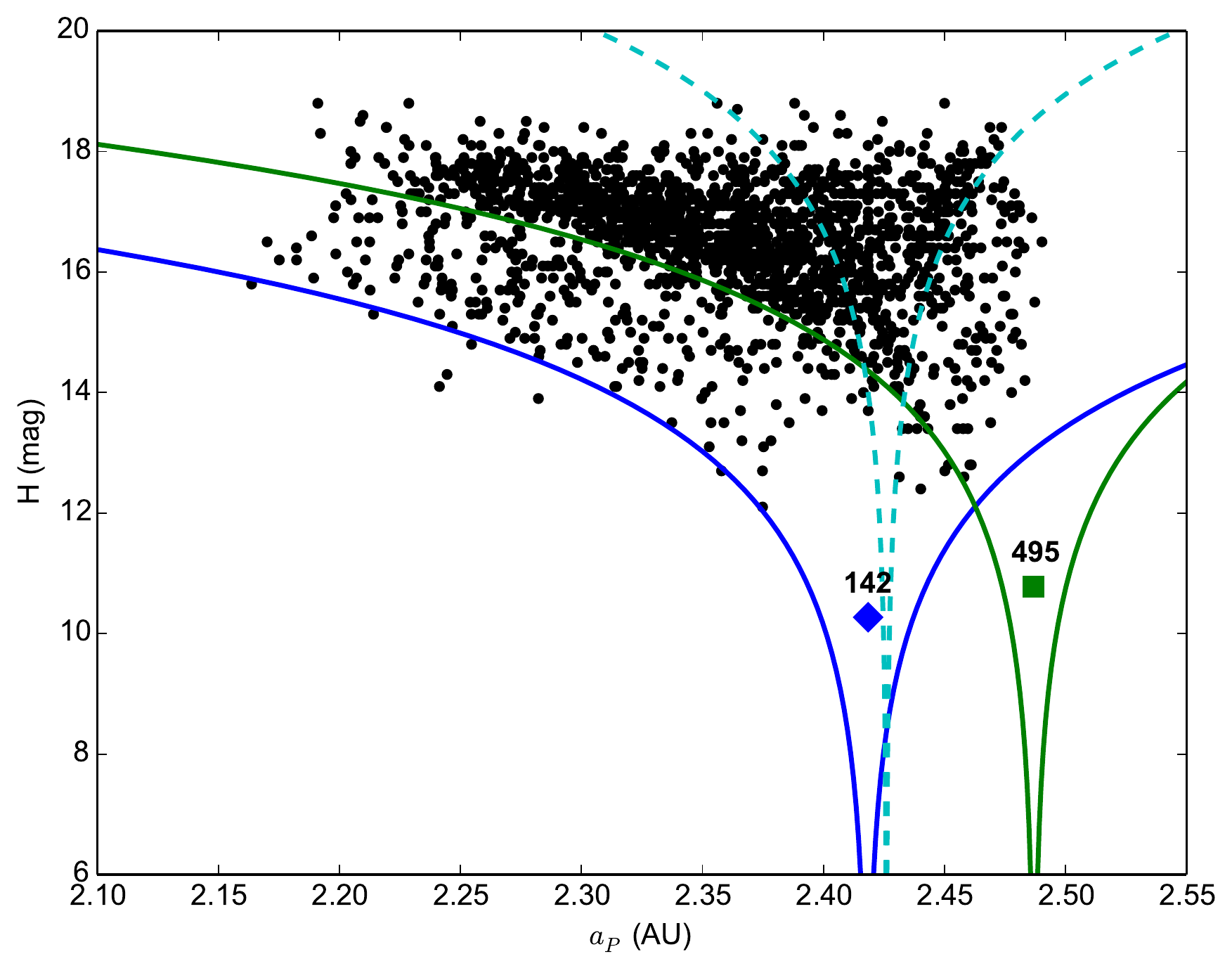}
\caption {\footnotesize The low-$a^*$ objects within the orbital range plotted in Fig.~\ref{11_low_astar} show at least two Yarkovsky V signatures, as noted by \citet{walsh2013} (cf. their Fig.~2). One structure is roughly centered on (142) Polana (blue diamond); the other is roughly centered on (495) Eulalia (green square). The curves correspond to $C = \pm0.169$~mAU (blue) and $C = \pm0.092$~mAU (green); the cyan lines show the Yarkovsky envelope for the Hertha2 family (cf. Fig.~\ref{12_hertha2_aH}), which is included in this low-$a^*$ population. }
\label{14_walsh}
\end{center}
\end{figure}

Other family structures among the dark asteroids in the Nysa-Polana region can be teased apart by considering the $a_\text{P}-H$ distribution in the context of the Yarkovsky evolution model, an approach utilized by \citet{walsh2013}. The $a_\text{P}-H$ distribution of the low-$a^*$ objects in the Nysa-Polana region is shown in Fig.~\ref{14_walsh}. This figure includes for context the cyan boundary $C$ parameter lines for the Hertha2 family (Fig.~\ref{12_hertha2_aH}), and plots new $C$ parameter lines for the two distinct families identified by \citet{walsh2013} (cf. Fig.~2 of that work), which are centered on (142) Polana and (495) Eulalia. These two families are visible in Fig.~\ref{14_walsh}; we refer to them as the Polana and Eulalia1 families, respectively. The latter is one of two structures that we identify with (495) Eulalia (the second is discussed below). 

\begin {center}
EDITOR: PLACE FIGURE \ref{14_walsh} HERE
\end {center}

Fig.~\ref{11_low_astar} demonstrates a phenomenon that was noticed first by \citet{walsh2013}: (495) Eulalia is significantly offset from its nominal family in $e_\text{P}$ ($e_\text{P} = 0.12$ vs. the family average of $e_\text{P} = 0.145$). \citet{walsh2013} modeled the long-term evolution of (495) Eulalia, and demonstrated that proximity to the Jupiter 3:1 resonance results in variability of the asteroid's proper eccentricity between 0.11 and 0.15 over 500 Myr. 

\begin{figure}
\begin{center}
\includegraphics [width=3.8in]{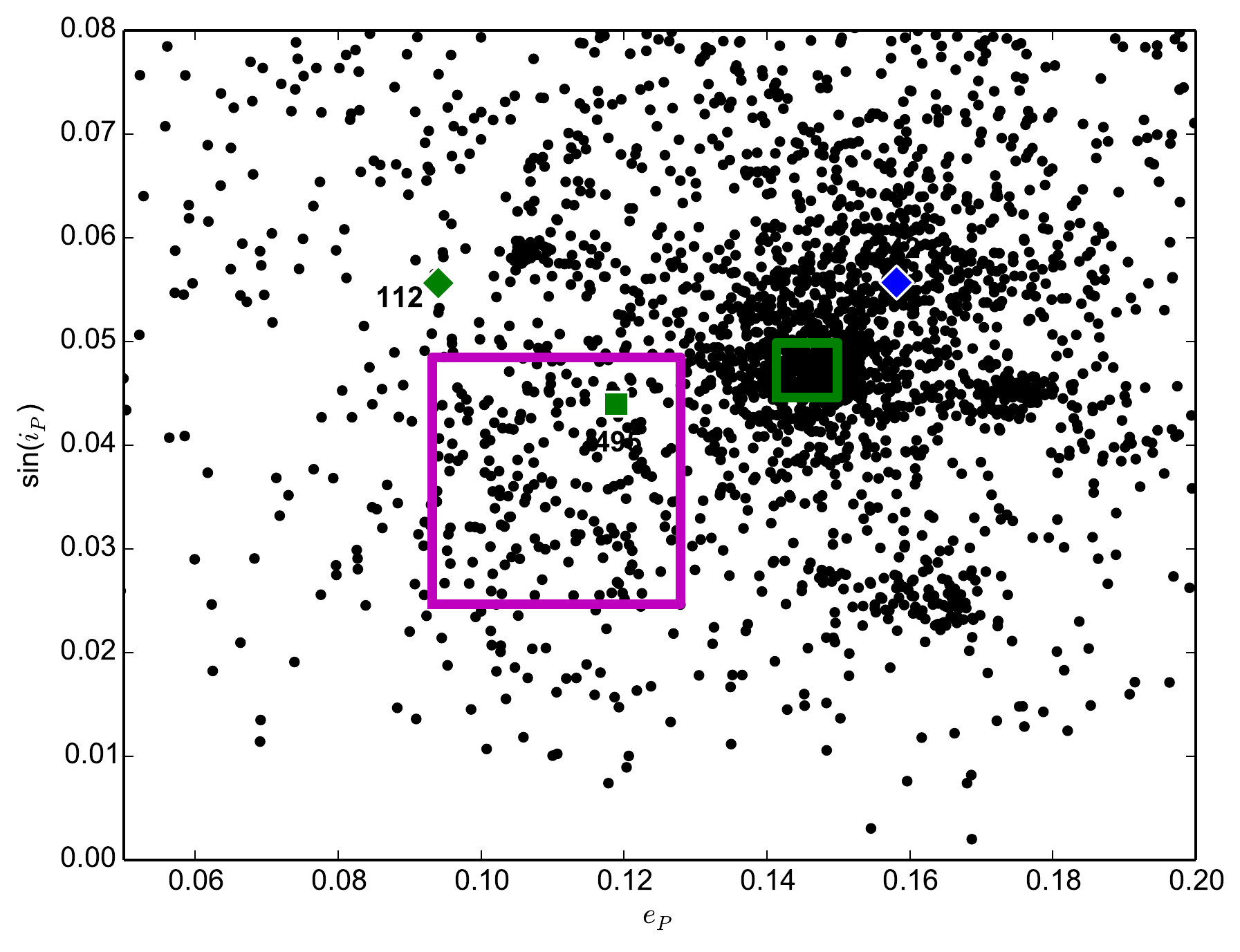}
\caption {\footnotesize Orbital distribution of low $a^*$ objects, similar to Fig.~\ref{11_low_astar}, but with adjusted ranges in $e_\text{P}$ and $\sin i_\text{P}$ to include the region near (495) Eulalia (green square). A diffuse cluster appears within the magenta box. \citet{walsh2013} associated this grouping with the object (112) Iphigenia (green diamond). The green box indicates the core of the Eulalia1 family, which makes up much of the population above the green line in Fig.~\ref{14_walsh}. }
\label{16_new_eulalia}
\end{center}
\end{figure}

In order to explore whether any other asteroids might be related to (495) Eulalia, we modify the range under consideration to $0.05 < e_\text{P} < 0.2$ and $0.0 < \sin i_\text{P} < 0.08$, and replot the distribution of low-$a^*$ objects in this new range (Fig.~\ref{16_new_eulalia}). A diffuse grouping appears near $e_\text{P} = 0.11$ and $\sin i_\text{P} = 0.035$, indicated roughly by the magenta box in Fig.~\ref{16_new_eulalia}. This grouping was noted by \citet{walsh2013} after removal of the dense Eulalia1 cluster, and was tentatively associated with (112) Iphigenia (indicated by the green diamond). 

\begin {center}
EDITOR: PLACE FIGURE \ref{16_new_eulalia} HERE
\end {center}

\begin{figure}
\vspace{-0.5in}
\begin{center}
  \begin{subfigure}{\textwidth}
    \begin{centering}
    \includegraphics [width=3.8in]{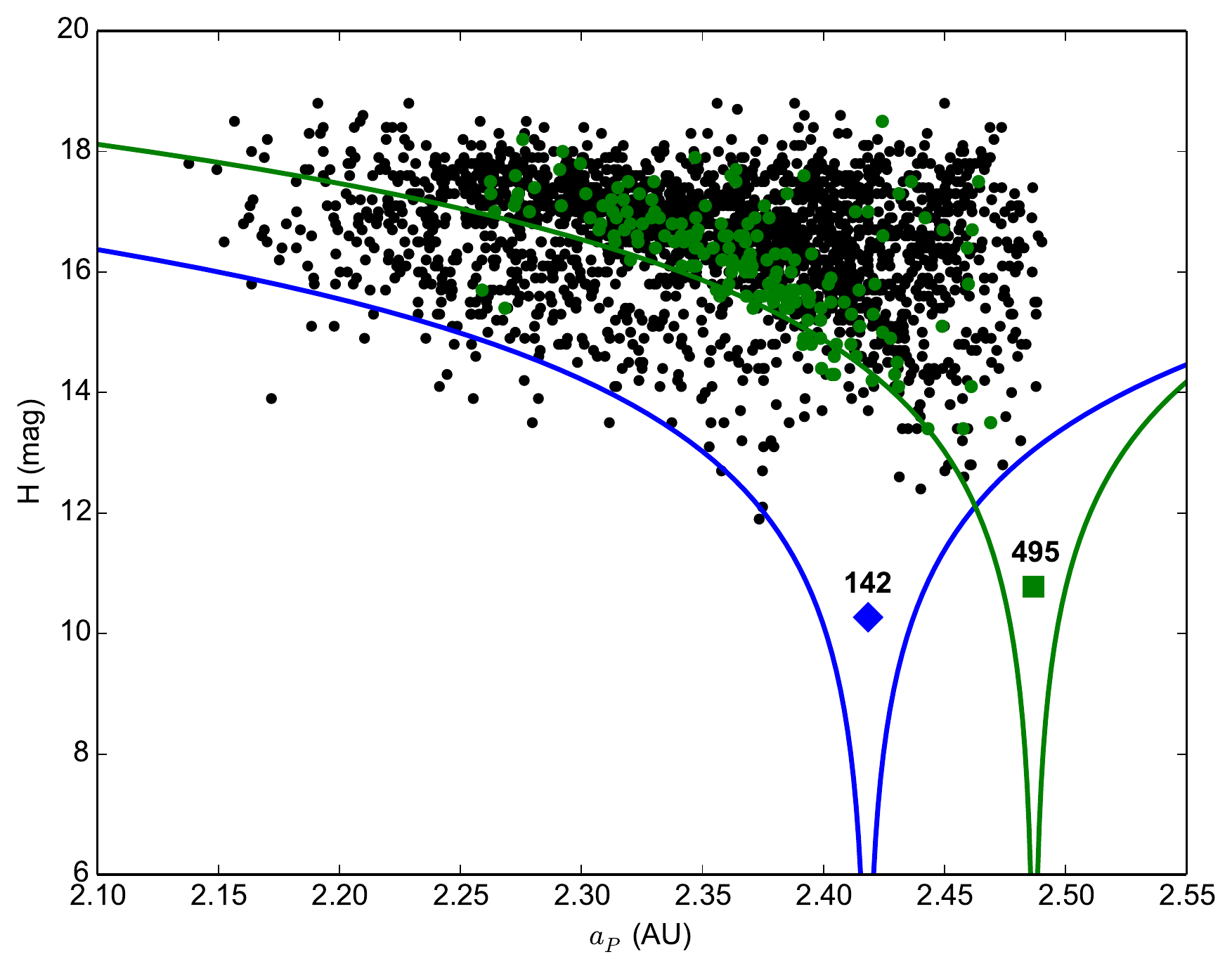}
    \vspace{-0.1in}
    \caption{}
    \label{17_eulalias_a}
    \end{centering}
  \end{subfigure}
  \begin{subfigure}{\textwidth}
    \begin{centering}
    \includegraphics [width=3.8in]{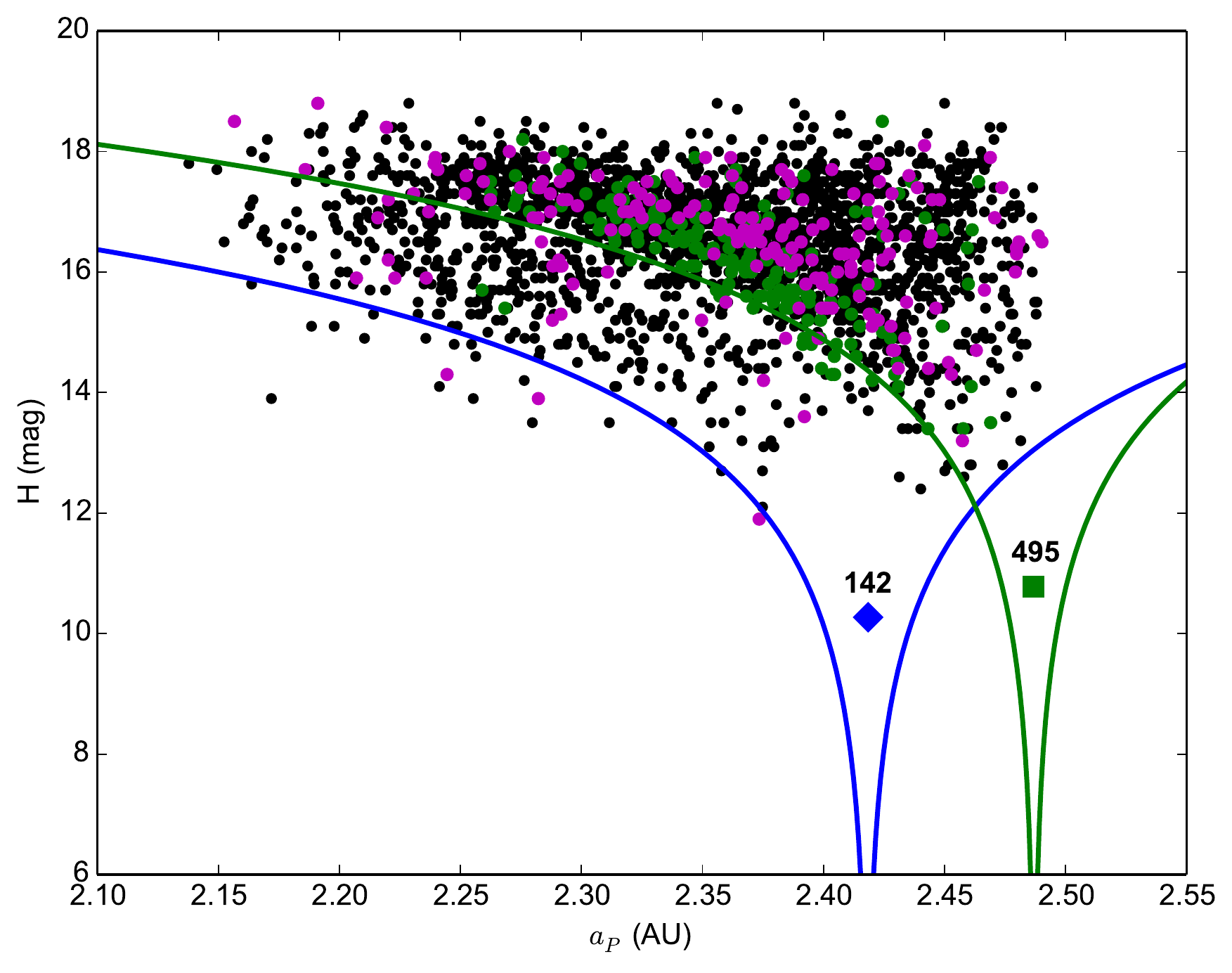}
    \vspace{-0.1in}
    \caption{}
    \label{17_eulalias_b}
    \end{centering}
  \end{subfigure}
  \caption {\footnotesize (a): The same as Fig.~\ref{14_walsh}, but with objects with orbits in the green box in Fig.~\ref{16_new_eulalia} (predominantly Eulalia1 members) highlighted in green. These objects are roughly centered on (495) Eulalia and bounded by the model curve of $C = -0.092$~mAU. (b): The same as (a), but with overplotted objects from within the magenta box in Fig.~\ref{16_new_eulalia}. These objects are roughly centered on (495) Eulalia as well, but do not extend all the way to the green curve, thus indicating that the structure within the magenta box in Fig.~\ref{16_new_eulalia} is distinct from the Eulalia1 family, and hence designated Eulalia2. }
\label{17_eulalias}
\end{center}
\end{figure}

Fig.~\ref{17_eulalias} shows the $a_\text{P}-H$ distribution of the objects in the green and magenta boxes in Fig.~\ref{16_new_eulalia}, that is, the objects of the Eulalia1 and the putative Iphigenia cluster. The latter objects show similar structure to the Eulalia1 objects, and can be described by the same Yarkovsky $C$ envelope. A preliminary best fit $C$ envelope for each is at $C = \pm0.075$~mAU (slightly different from that found by \citet{walsh2013}, $C = \pm0.092$~mAU). Note that the two groups display a distinct offset in semimajor axes ($\delta a_\text{P} \sim 0.035$~AU); this offset, coupled with the identical boundary $C$ envelopes for the two clusters, leads us to designate the second cluster as ``Eulalia2'' and interpret it as possible evidence for an anisotropic ejection field resulting from the Eulalia1 collision event, rather than associate it with (112) Iphigenia. The implications of the structures of these two clusters are discussed further in Section \ref{discussion_eulalias}.

\begin {center}
EDITOR: PLACE FIGURE \ref{17_eulalias} HERE
\end {center}

The semimajor axis offset between these two populations was not discussed by \citet{walsh2013}, and indeed is not apparent in the corresponding figure in that work (their Fig.~20). Walsh et al. relied on the hierarchical clustering method (HCM) to identify groupings of objects for comparison. While the HCM is ideal for identifying clusters in multidimensional space, it is less ideal for comparison of neighboring families, as it tends to include in each sample objects from the neighboring family or families, which muddies the comparison. Instead, we select for comparison two distinct sets of objects (within the green and magenta boxes in Fig.~\ref{16_new_eulalia}) for which we expect minimal contamination between sets, which enables us to identify the small offset in $a_\text{P}$ among the Eulalia1 and Eulalia2 structures.

\section{Discussion}
\label{discussion}

We have identified five distinct clusters of asteroids within what has generally been called the Nysa-Polana orbital region of the asteroid belt. Two of these clusters appear to be associated with collisions on (135) Hertha; two other clusters appear to be associated with (495) Eulalia; one is associated with (142) Polana. None of these is associated with (44) Nysa.

This refined interpretation of the dynamical structure of the Nysa-Polana region has been made possible by the availability of color and albedo information from the SDSS and WISE projects, which have increased the dimensionality of data available for cluster analyses beyond simply the proper orbital element distribution. The analysis of the clusters within the Nysa-Polana region depends particularly upon this increased dimensionality, due to the overlap within the region of multiple large families with varied reflectance properties.

Other recent work has explored clusters within this newly extended parameter space using an automated approach based on the hierarchical clustering method \citep{masiero2013,milani2014,carruba2013}. Our approach differs from theirs in that we (like \citet{walsh2013}) have also incorporated the effects of Yarkovsky semimajor axis drift, which results in recognizable patterns in the relationship between semimajor axis spread and asteroid size (equivalently, the absolute magnitude), specifically the characteristic V shape in $a_\text{P}$ vs. $H$. In addition, we use the slope of the $a_\text{P}-e_\text{P}$ correlation among the Hertha1 family members as a constraint on the orbit configuration of the parent object at the time of the collision, enabling an improved evaluation of the Yarkovsky semimajor axis dispersion of the family independent of the collisional dispersion.

\subsection{Hertha1 and Hertha2}
\label{discussion_herthas}

The collision that formed the Hertha1 family $300^{+60}_{-50}$~Ma (Section \ref{high_astar}) distributed ejecta across nearly the entire span of the inner main belt in semimajor axis, depositing material directly into Mars-crossing orbits at the time of the event. Under the influence of the Yarkovsky effect, the Hertha1 family has since provided a steady stream of material of various sizes into both the Mars-crossing region and the Jupiter 3:1 resonance. Hence, it is plausible that the Hertha1 family has been a source population for significant impact events of various magnitudes on Earth in the past 300 million years, a time span which includes the event hypothesized to be connected with the Permian-Triassic extinction \citep[251.4 Ma,]{raup1982,jin2000}.

The weak Mars 1:2 resonance affects the semimajor axis and eccentricity distribution of the Hertha1 family \citep{gallardo2011}. The effect of this resonance is most noticeable in the dearth of objects in a depleted zone with $a_\text{P} < 2.43$~AU and $e_\text{P} > 0.195$ in Fig.~\ref{04_ae}, compared for example with the density of objects just below ($e_\text{P} < 0.19$) in the same figure. When objects drifting from the family center toward higher semimajor axes reach the resonance, it diffuses their eccentricities, resulting in higher average eccentricity of the resonant population. Thus, for objects starting at low eccentricities, the boost moves them to a higher-$e_\text{P}$ section of the Hertha1 family; for objects starting at high eccentricities, the boost may remove them from the family altogether, hence the paucity of blue dots in the depleted zone. This paucity demonstrates also the effects of the combined Yarkovsky and YORP effects on the Hertha1 family; objects that would otherwise have drifted back and forth as a result of YORP cycles, continually re-filling the depleted zone, have instead been removed from the family to higher eccentricities.

The presumed parentage and nomenclature of the Hertha1 family has varied throughout the literature (Table~\ref{table1}). Unlike the small, low-$a^*$ Hertha2 family, whose objects show reflectance properties consistent with X-type spectra and thus similar to (135) Hertha itself, the large Hertha1 family shows average reflectance properties consistent with S-type spectra. S-type objects would not generally be expected to originate from an X-type parent body, given the standard picture of S-type objects originating from chondritic, undifferentiated parent objects. However, \citet{weiss2013} suggested that some chondritic bodies may originate from the unmelted crusts of partly differentiated planetesimals. Moreover, \citet{thomas2014} have suggested that the Massalia family may have olivine-rich objects among its predominantly S-type members. Accordingly, it is plausible that the X-type (135) Hertha is associated with the S-type Hertha1 family.

In the following subsections, we consider four scenarios for the origin of the Hertha1 and Hertha2 families that might explain their different reflectance properties. None of the scenarios explain the observations completely; therefore the actual origin scenario may be a combination of these or other effects.

\subsubsection{Scenario 1: Unrelated families}
\label{unrelated}

The possibility that both Hertha1 and Hertha2 originated from two unrelated collisions on different parent objects, one of which (Hertha1) was catastrophically disrupted, cannot be ruled out here. One approach to understanding the type of collision expected to have formed the Hertha1 family involves comparison of the observed size-frequency distribution (SFD) of the family members with models of SFDs produced by catastrophic and cratering collision events. However, the incompleteness of the reflectance property databases, as well as the depletion of primarily smaller objects due to the nearby large resonances, have introduced significant bias into the family's observed SFD. Future efforts will explore the connection between the Hertha1 family and (135) Hertha in detail, including numerical integrations to determine the possible evolution of the proper orbital elements of (135) Hertha \citep[similar to the method used for (495) Eulalia by][]{walsh2013}. In the meantime, the close dynamical proximity of both Hertha1 and Hertha2 families to the asteroid (135) Hertha remains strong evidence that they share this common parent, and warrants discussion of additional origin scenarios.

\subsubsection{Scenario 2: Single collision on a partially differentiated parent}
\label{core}

A single collision on a partially differentiated planetesimal could produce, in principle, a family of objects with varied material properties. The tighter Yarkovsky V of the Hertha2 family ($C = \pm0.012$~mAU vs. $C = \pm0.045$~mAU for the Hertha1 family) would appear to suggest a younger age, and hence a separate collision. However, a single collision could still explain the two different Yarkovsky V's if the tighter V reflects a decreased Yarkovsky semimajor axis drift rate instead of a younger age, with the objects of the Hertha2 family drifting outward more slowly after the original collision.

A slower drift rate for the Hertha2 objects could result from their lower average albedo, $p_\text{V}$. The small numbers and small sizes of the Hertha2 family objects complicate the determination of the average albedo; however, it would have to be $\sim 0.014$ to account for the reduced semimajor axis spread, which is implausibly low (asteroid (135) Hertha has $p_\text{V} = 0.15$). Thus, the relatively narrow V of the Hertha2 family cannot be accounted for by lower albedo.

A higher average bulk density might alternatively explain the slow drift. The Yarkovsky drift rates scale inversely with density, so to account for the different V shape relative to the Hertha1 S-type family (with $\rho_\text{S} = 2.7$~g/cm$^3$), the average density that would be required for the Hertha2 objects would need to be 8.6~g/cm$^3$. For comparison, the bulk density of Hertha is $\rho_\text{H} = 5.23$~g/cm$^3$, and the average for the $\text{X}_\text{k}$-type objects is $\rho_{\text{X}_\text{k}} = 3.8$~g/cm$^3$ \citep{carry2012}. The larger density could be achieved if the family were composed primarily of metal objects; however, the family members' apparent association with the lower-density (135) Hertha does not imply that this is the case. Thus, density differences probably cannot explain the reduced spread of the Hertha2 family, and a more recent formation event seems the most plausible explanation.

Further evidence for a separate formation event comes from the observed $a_\text{P}-e_\text{P}$ correlation among the Hertha1 family members' orbits (Section \ref{03_high_astar}), and the lack of one among the Hertha2 family members. If the Hertha2 members formed in the same collision as the Hertha1 members, they would be expected to share this correlation. Thus the most likely explanation is that two separate collision events occurred.

\subsubsection{Scenario 3: Two collisions on a partially differentiated parent}
\label{subsequent}

A series of two collisions occurring on a partially differentiated parent could excavate families with varied material properties from different regions in the parent body. In the case of the Hertha families, an initial collision into a parent body's unmelted, chondritic crust could form the S-type Hertha1 family, leaving a large X-type remnant, (135) Hertha. A subsequent impact onto (135) Hertha could then create the X-type Hertha2 family.

A collision of magnitude sufficient to create the Hertha1 family might be expected to produce more than one remnant Hertha-like X-type among its fragments. The nearest object with reflectance properties similar to (135) Hertha is the large E-type (44) Nysa, which could have conceivably come from the original Hertha1 impact, although it is significantly displaced from the family in proper inclination. \citet{zellner1977} and \citet{kelley1994} attempted to link (44) Nysa and (135) Hertha in various differentiation structures, and \citet{rivkin1995,rivkin2000} demonstrated that the spectra of both objects share a hydration feature that could suggest a common origin. Nevertheless, the issue of whether a first impact could have cleanly removed the S-type material leaving only one X-type remnant for subsequent impact remains uncertain.

\subsubsection{Scenario 4: Two collisions with shock darkening or space weathering effects}
\label{shock_darkening}

The effects of shock darkening (through the presence of impact melt) can modify the apparent reflectance properties of asteroids without changing their basic material properties \citep{britt1994}. In the case of the Hertha families, the impact that formed the Hertha1 family could produce remnants such as (135) Hertha with a significant amount of shock-darkened material, resulting in the later excavation of material with X-type appearance during the Hertha2 formation event. 

\citet{reddy2014} demonstrated the effects of shock darkening among laboratory samples of LL chondrite material mixed with impact melt, and matched the resulting spectra with observations of the X-type Baptistina family members. In appearance, the difference in $a^*$ color between the Hertha1 and Hertha2 families ($-0.07$ vs. 0.13) is strikingly analogous to that of the nearby Flora and Baptistina populations: The Baptistina members lie in nearly the same place in $a^*$ and $i-z$ color space as the Hertha2 members, as seen by comparison of the location of the Hertha2 family in Fig.~\ref{13_hertha_region} with the location of the Baptistina family in a similar plot \citep[e.g., Figs. 3e and 3f of][]{dykhuis2014}; and both the Flora and Hertha1 populations show reflectance properties typical of S-type objects.

Given the results of \citet{reddy2014} for the Baptistinas, it is worth considering the possibility that the X-type appearance of (135) Hertha and the Hertha2 family members results from shock darkening, i.e. the presence of significant impact melt matrix surrounding fragments of the original S-type material. In this case, a plausible origin hypothesis is that the impact that formed the Hertha1 family left behind a shock-darkened fragment, (135) Hertha, which experienced a later collision that formed the darkened Hertha2 family. 

Another explanation for the difference in $a^*$ between the two Hertha families could be the effects of space weathering, which raises the $a^*$ value of exposed asteroid surfaces over time, resulting in higher $a^*$ for older families. In this case, the older Hertha1 family would be expected to show higher average $a^*$ values than the Hertha2 family, as observed.

Consideration of the surface properties of the four subfamilies of the large, S-type Koronis family \citep{molnar2011} raises issues regarding both the shock-darkening and the space-weathering hypotheses. These subfamilies range in age from 5.8~My to 300~My, and display a linear trend in $a^*$ from 0.02 for the youngest family to 0.07 for the oldest \citep{molnar2011,dykhuis2014}. This range of $a^*$ values is consistent with the color change expected as a result of space weathering, based on a study of various other families by \citet{willman2010}. The difference in $a^*$ between the two Hertha families is four times greater, so space weathering is unlikely to be the key factor in explaining this difference. If, instead, shock darkening is the dominant factor in the different appearance of the Hertha families, it raises the question of why similar effects have not led to more diversity among the Koronis families. The answer may simply lie in the unknown specific details of the collisions responsible for each family group.

\subsection{Eulalia1 and Eulalia2}
\label{discussion_eulalias}

The low-albedo families in the Nysa-Polana region are thought to be the most likely source families for asteroid (101955) Bennu, the target of the OSIRIS-REx sample return mission \citep{campins2010,walsh2013,bottke2014}. \citet{bottke2014} determined the probability that Bennu originated from what we call the Polana family to be 70\%, with the probability of an origin from what we call the Eulalia1 family of 30\%. The Eulalia2 objects identified here have similar orbital and reflectance properties, and might also be a candidate source population for Bennu. The possibility of a collisional connection between the Eulalia1 and Eulalia2 clusters, discussed below, could help further constrain the dynamics of the Bennu source region.

The Eulalia1 and Eulalia2 clusters show very similar structure in $a_\text{P}-H$ space, and can be roughly fit with $C$ envelopes that match, except for a semimajor axis offset of about $\delta a_\text{P} \sim 0.035$~AU (Section \ref{other_dark}, Fig.~\ref{17_eulalias}), with Eulalia1 centered at 2.46~AU and Eulalia2 centered at 2.495~AU. These values bracket the current $a_\text{P}$ value for (495) Eulalia ($a_\text{P}$ = 2.4867~AU). The Eulalia1 and Eulalia2 clusters also bracket (495) Eulalia in $e_\text{P}$ and $\sin i_\text{P}$ (Fig.~\ref{16_new_eulalia}). \citet{walsh2013} noted that the long-term evolution of (495) Eulalia ranges between $e_\text{P}$ values of 0.11 to 0.15; this range is contained within the larger range of the combined Eulalia1 and Eulalia2 families.

In light of these dynamical coincidences (matching $C$ envelopes, bracketing in each orbital dimension), it is plausible that the Eulalia1 and Eulalia2 clusters were originally part of the same collisional family, centered near the current position of (495) Eulalia. In this case, the twin ``lobes'' of the family protruding in opposite directions in $a_\text{P}, e_\text{P}$ and $\sin i_\text{P}$ could be explained by a non-isotropic ejecta pattern. In this scenario, due to the parent object's proximity to the Jupiter 3:1 resonance, a significant number of the original objects would have been deposited directly into the resonance (though not beyond it, see the discussion in Section 2.5 of \citet{walsh2013}). The resonance would then have dramatically reduced the number of objects in the high-$a_\text{P}$, low-$e_\text{P}$, low-$\sin i_\text{P}$ lobe of the family. Subsequent Yarkovsky drift would then have removed another half of the remaining family (the prograde objects) via drift into the Jupiter 3:1.

The above scenario could be tested dynamically, by modeling the best fit values of the $C$ boundaries and $a_\text{P,0}$ for the Eulalia1 and Eulalia2 clusters to confirm the similarity of $C$ for the two and determine their offset in semimajor axis. This modeling is necessary to confirm the significance of the observed offset, as the sample of Eulalia2 members (from the magenta box in Fig.~\ref{16_new_eulalia}) is fairly diffuse and thus might be expected to contain a significant number of background objects. In addition, orbital integrators that incorporate the Yarkovsky and YORP effects could be used to simulate and reproduce the observed $a_\text{P}-H$ distributions of the family from a single collision.

Lastly, spectral comparison of the reflectance properties of the Eulalia1 and Eulalia2 structures can also test the above origin scenario. The SDSS and WISE databases provide reflectance data for a large number of asteroids in the region; however, the completeness of these databases is only around 20\% and 10\%, respectively. Future surveys such as the Gaia space mission of the European Space Agency\footnote{sci.esa.int/gaia/} and the Large Synoptic Survey Telescope (LSST)\footnote{www.lsst.org/lsst/} are expected to significantly enhance these databases in the near future, providing opportunities for further analysis. The Polana and Eulalia1 and Eulalia2 structures are all similar in the wavelength range available to the SDSS, and a survey of 72 objects in the region shows spectral homogeneity in the NIR wavelengths as well (N. Pinilla-Alonso, in preparation). Future work in this area might possibly distinguish or further connect the Eulalia1 and Eulalia2 structures.

\section* {ACKNOWLEDGMENTS}

{\scriptsize The authors are grateful to Miroslav Broz and an anonymous reviewer for their comments, which significantly improved the quality of this manuscript. We acknowledge the use of data from the Sloan Digital Sky Survey, and thank the Sloan team and its sponsors (see http://www.sdss.org). In addition, this publication makes use of data products from the Wide-field Infrared Survey Explorer and NEOWISE, which is a joint project of the University of California, Los Angeles, and the Jet Propulsion Laboratory/California Institute of Technology, funded by the National Aeronautics and Space Administration. We thank Bill Bottke, Dante Lauretta and the OSIRIS-REx science team for their guidance and advice. In addition, we thank Nathaniel Dykhuis for his assistance with the statistical analysis. MJD was supported by an NSF Graduate Research Fellowship, Award No. DGE-1143953, with additional support from the OSIRIS-REx Dynamical Evolution Working Group. 

\bibliographystyle{plainnat}
\bibliography{bibliography}
}

\end{document}

%% file: Tables/summary_table_tolatex.txt
\begin{table}[H]
{\scriptsize
\begin{tabular}{ | l | c | c c c c c c c c | }
 \hline 
 This work   & Class   & Ze77   & Wi79   & B89    & Za95   & C01     & MD05           & Wa13       & M14 \\ [1.0ex]
\hline
 Hertha1     & S       & Hertha & W-160  & Hertha & Nysa   & Mildred & Mildred/Hertha & -          & Burdett \\
 Hertha2     & X       & -      & -      & -      & -      & -       & Hertha         & -          & -       \\
 Polana      & C       & Nysa   & W-24   & Polana & Polana & Polana  & McCuskey       & New Polana & Polana  \\
 Eulalia1    & C       & "      & "      & "      & "      & "       & "              & Eulalia    & -       \\
 Eulalia2    & C       & -      & -      & -      & -      & -       & -              & Iphigenia? & -       \\
\hline
\end{tabular}
\\
}